\documentclass[aps,prx,superscriptaddress,floatfix,showpacs,notitlepage,twocolumn, 10pt]{revtex4-2}

\usepackage{graphicx}
\usepackage{pifont}
\usepackage{longtable}
\usepackage{dcolumn}
\usepackage{bm}
\usepackage{amsmath,amssymb}
\usepackage{amsthm}
\usepackage{mathrsfs}
\usepackage{indentfirst}
\usepackage{float}
\usepackage{braket}
\usepackage[utf8]{inputenc}
\usepackage{bm}
\usepackage{graphicx}
\usepackage{tikz}
\usepackage{physics}
\usetikzlibrary{arrows, shapes.gates.logic.US, calc}
\usetikzlibrary{angles,quotes}
\tikzstyle{branch}=[fill, shape=circle, minimum size=3pt, inner sep=0pt]
\usepackage{blochsphere}
\usepackage{mathtools}
\usepackage{color,graphicx} 

\usepackage{bbold}
\usetikzlibrary{quantikz}
\usepackage{xcolor}

\usepackage{hyperref}
\hypersetup{
 colorlinks=true,
 citecolor=black,
 linkcolor=blue,
 urlcolor=blue,
 pdfpagemode=UseNone,
 pdfstartview=FitH}

\usepackage{subfiles}
\usepackage{subfigure}

\newcommand{\rr}[1]{\mathrm{#1}}
\newcommand{\cl}[1]{\mathcal{#1}}

\definecolor{deeppink}{RGB}{255,20,147}
\definecolor{GCyan}{RGB}{23, 190, 207}
\definecolor{orange}{RGB}{255,165,0}
\definecolor{tabgreen}{rgb}{0.137, 0.545, 0.270}

\footnotetext{These authors contributed equally to this work.}

\begin{document}

\preprint{APS/123-QED}
\title{Hierarchical generation and design of tree-codes for resource-efficient loss-tolerant quantum communications}

\author{Francesco Cesa$^\star$}\email{francesco.cesa@phd.units.it}
\affiliation{Department of Physics, University of Trieste, Strada Costiera 11, 34151 Trieste, Italy}
\affiliation{Istituto Nazionale di Fisica Nucleare, Trieste Section, Via Valerio 2, 34127 Trieste, Italy}
\affiliation{Institute for Quantum Optics and Quantum Information of the Austrian Academy of Sciences, 6020 Innsbruck, Austria}
\affiliation{Institute for Theoretical Physics, University of Innsbruck, 6020 Innsbruck, Austria}
\author{Tommaso Feri$^\star$}\email{tommaso.feri@phd.units.it}
\affiliation{Department of Physics, University of Trieste, Strada Costiera 11, 34151 Trieste, Italy}
\affiliation{Istituto Nazionale di Fisica Nucleare, Trieste Section, Via Valerio 2, 34127 Trieste, Italy}
\author{Angelo Bassi}
\affiliation{Department of Physics, University of Trieste, Strada Costiera 11, 34151 Trieste, Italy}
\affiliation{Istituto Nazionale di Fisica Nucleare, Trieste Section, Via Valerio 2, 34127 Trieste, Italy}

\begin{abstract}
We develop novel protocols for generating loss-tolerant quantum tree-codes; these are designed to safeguard information against qubit losses, with wide applications in quantum communications. Contrary to previous proposals, our method enables \emph{top-to-bottom} fast encoding and decoding, thereby reducing losses due to the lagging and photon-reordering at the repeater stations. At the hardware level, we show how to achieve this with a single quantum emitter equipped with a \emph{static} feedback mechanism, which we leverage to engineer entangling gates between a fed-back qubit and multiple emitted qubits in parallel. In addition, analyzing typical patterns within the error-correction decoding graphs, we find optimizations of the structure of tree-codes, which enable improved performance by also reducing the code size; these are based on the introduction of \emph{asymmetries} in the code, which mimic the intrinsic adaptiveness of the recovery procedure. We show numerically that these improvements together significantly enhance the loss-correction performance. Specifically, focusing on quantum repeater protocols, we show that our fast recovery scheme (decoding-encoding) allows for improved repeater rates with smaller photon numbers per code.

\end{abstract}
\maketitle

\section{Introduction}
Qubit loss is a primary source of errors when photons are employed in quantum information protocols~\cite{o2009photonic, couteau2023applications}. Particularly, in long-range quantum communications, e.g. for quantum key-distribution (QKD)~\cite{bennett2014quantum, ekert1991quantum, scarani2009security, portmann2022security}, transmissions are severely limited by the fact that photon losses grow exponentially with the length of the channel. Indeed, while experiments have successfully demonstrated quantum networking at remarkable distances~\cite{entanglement_distrib_Nature_2024, metropolitan_networ_Nature_2024,stolk2024metropolitan, neumann2022continuous,QKDexp_4660_Nature_2021,entanglement_distrib_nature_comm_2022, ribezzo2023deploying}, further advances crucially depend on overcoming losses~\cite{Pirandola_2017, wehner2018quantum}.\\
\indent In principle, losses can be faced by employing quantum repeater schemes~\cite{briegel1998quantum,azuma2015all,HANNES_PRX, azuma2023quantum} and encoding the information in entangled loss-tolerant codes (LTCs), i.e. quantum states where the information is stored non-locally in correlations among several physical photonic qubits~\cite{VARNAVA_LOSS, bell2023optimizing}; then, the information can be restored even in presence of significant losses among the components of the LTC. This can also be understood as a form of quantum error-correction (QEC), where errors are in the form of \emph{erasures} - i.e., their location is known to the decoder~\cite{grassl1997codes}. These QEC settings feature much higher thresholds than more general schemes: for instance, it is known that losses can be corrected efficiently (i.e., with polylogarithmic resource overheads) up to the break-even point of a $50\%$ error-rate~\cite{VARNAVA_LOSS}. However, in practice the possibility of correcting photon losses is currently limited by the hardness of efficiently generating entangled photonic states in the first place, posing a bold question mark on the feasibility of photonic quantum information processing at large scales. Correspondingly, in recent years the development of methods for the efficient generation and error-correction of LTCs has become a topic of increasing interest~\cite{economou_prx,HANNES_PRX,ZHAN_PRL,bell2023optimizing, wo2023resource, Hilaire_2023,Aqua_2025}, unveiling key technical challenges both on the theoretical and on the experimental side~\cite{ECONOMOU_PERFORMANCE}. \\
\begin{figure*}[t!]
    \begin{minipage}{1.0\textwidth}
        \includegraphics[width=\textwidth]{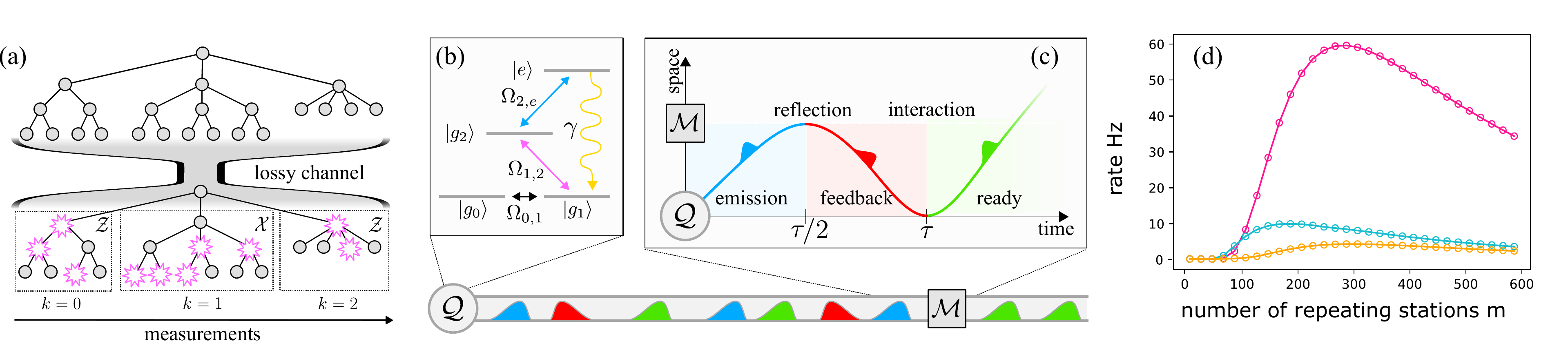}
    \end{minipage}%
\caption{(a) The recovery protocol for a generic loss-tolerant quantum code after it undergoes a lossy channel: we stress the adaptiveness of the destructive measurements, which in the figure are performed from the left side to the right side.  (b,c) A quantum emitter $\cl{Q}$, with the level structure shown in the left inbox, is coupled to a waveguide, and can be driven in such a way to deterministically emit single photons. Moreover the photons, after the emission, are fed-back to $\cl{Q}$ once: in the figure, we represent this through the component $\cl{M}$, which can be realized e.g. with a switchable mirror. We color in blue photons which have been emitted but still must be reflected and fed-back, in red those which have been reflected and are feeding-back, and in green those which have already interacted with $\cl{Q}$. The trajectory of one photon is also shown in the space-time diagram in the right inbox, where $\tau$ is the time-delay between emission and feedback. (d) The rate of Eq.~\ref{rate_formal}, for $L = 300\mathrm{km}$, as a function of the number $m$ of repeater stations disposed along the transmission line. The markers correspond to:  {\protect\tikz \protect\draw[draw=deeppink, thick] (0,0) circle (0.3em) -- (-0.4,0) -- (0.4,0); } top-to-bottom (hierarchical) generation with a single emitter of asymmetric LTCs, as proposed in this work; {\protect\tikz \protect\draw[draw=orange, thick] (0,0) circle (0.3em) -- (-0.4,0) -- (0.4,0) ;} bottom-to-top generation of symmetric LTCs as proposed in Refs.~\cite{HANNES_PRX, ECONOMOU_PERFORMANCE}, with maximum branching parameter of the tree-graph set to $5$;  {\protect\tikz \protect\draw[draw=GCyan, thick] (0,0) circle (0.3em) -- (-0.4,0) -- (0.4,0) ;} the same bottom-to-top method, but  without restrictions on the number of children qubits.}
\label{fig_1}
\end{figure*}
\indent In this work, we contribute to the feasibility and performance of loss-correction in two main directions. First, we develop novel protocols for the \emph{generation} of photonic LTCs, specifically from the family of tree-graph codes~\cite{VARNAVA_LOSS, HANNES_PRX}. These protocols advance existing methods (also using similar hardware~\cite{ZHAN_PRL}) by allowing for a \emph{hierarchical} generation of tree LTCs - a feature that, although recognized to be highly desirable~\cite{ECONOMOU_PERFORMANCE}, had previously been achieved only at the cost of substantial overheads~\cite{li2022photonic}. In line with expectations, our numerical analyses show that our hierarchically generated LTCs feature improved loss-tolerance properties. Second, we tackle the design of the LTCs themselves, for which we introduce a refined design approach, which we motivate both analytically and numerically: in contrast with current proposals, we show that the introduction of judicious asymmetries in the LTCs can systematically improve the success-rates by also lowering the required photon number. \\
\indent Our methods are motivated and inspired by many recent advancements in the deterministic coupling of waveguides with controllable, qubit-like quantum systems across various platforms, opening the door for novel directions in quantum networking~\cite{wehner2018quantum,borregaard2019quantumnetworksdeterministicspinphoton}, where entanglement between single photons is mediated deterministically by programmable emitters~\cite{uppu2021quantum, lodahl2015interfacing, le2022dynamical, reuer2022realization, ferreira2024deterministic, Thomas_2024, meng2023photonicfusionentangledresource,Huet_2025}. In this spirit, our scheme is deterministic and only requires one single quantum emitter; this provides practical simplifications to the system layout, and is aligned with current and near-term experimental capabilities. Altogether, our results go in the direction of improving the performance (i.e., the loss-tolerance), by simplifying the physical resources (number of photons and of `matter qubits'). \\
\indent We show that several physical implementations benefit of our protocols, including silicon-vacancy (SiV) color centers~\cite{Silicon_vacancy_coherece,bhaskar2020experimental}, semiconductor quantum dots~\cite{tcoh_qdot,gamma_qdot} and neutral atoms~\cite{tcoh_atomo,gamma_atomo}. We note that our results have remarkable consequences for atoms in cavities, which had previously been ruled out for the implementation of loss-tolerant protocols similar to the ones studied here~\cite{ECONOMOU_PERFORMANCE}, even though featuring high fidelity operations and reliable interfaces with waveguides. In contrast, we find that the methods introduced here conspire constructively to allow high quality loss correction also in these setups. In fact, we show that our reductions in physical requirements combine favorably with outstanding parameters of neutral atom settings (such as very long coherence times), to overcome their main limitations (e.g., slow emission rates). This eventually results in such systems surpassing the threshold for secure loss-tolerant communication, showcasing a clear example of the practical improvement allowed by our constructions.\\
\indent This work is organized as follows. In the next section (Section~\ref{section2}), we start by introducing the main figures of merit considered, and we highlight a series of relevant aspects. Then, we give a broad overview of the ideas and results, providing an intuition in the main insights and methods. In Section~\ref{section3}, we formally introduce the LTCs we are interested in. In Section~\ref{section4}, we present our generation scheme, providing detailed account of the considered experimental paradigm and end-to-end protocols for all the steps involved. In Section~\ref{section5}, we present our code optimization strategy, based on asymmetric LTCs. In Section~\ref{section6} we analyze numerically the presented methods on the explicit example of one-way quantum repeaters; in Section~\ref{hardware considerations} we discuss physical realizations with present and near-term technology. Finally, in the last part (Section~\ref{conclusion}) we discuss and summarize the contents of our work.

\section{Main concepts and results}\label{section2}

For concreteness, we will benchmark our methods on the application of quantum communications, where losses have the most dramatic impact. The underlying main-picture is therefore the one of quantum repeater schemes~\cite{azuma2015all, HANNES_PRX, azuma2023quantum}: information is not only encoded redundantly in LTCs, but also the transmission line is divided in smaller portions, and the signal is periodically restored (decoding and re-encoding) at intermediate repeater stations. Thus, the main figure of merit we are eventually interested in is a \emph{communication rate} of the generic form
\begin{equation}\label{rate_informal}
    \mathcal{R} = \mathcal{R} \left( L, m, \mathscr{S}_\text{phys} ;  \;\mathscr{E}_\text{QEC}, \mathscr{P}_\text{gen}  \right).
\end{equation}
This naively depends on the overall transmission distance $L$, on the number of repeater stations $m$, and on the physical system $\mathscr{S}_\text{phys}$ employed to operate at the various stations; in addition, $\mathcal{R}$ depends strongly also on the abstract encoding $\mathscr{E}_\text{QEC}$ for QEC, i.e. the LTC, and on the protocols $\mathscr{P}_\text{gen}$ employed for executing each step, e.g. for the preparation of the LTC. Here, our principal goal is to design $\left(\mathscr{E}_\text{QEC},\mathscr{P}_\text{gen}\right)$, so to reach high rates $\mathcal{R}$ over long distances $L$, with minimal resources (e.g., minimal repeater stations $m$, or small error-correcting codes).\\
\indent More specifically, we envision a final goal of implementing quantum key-distribution (QKD), and correspondingly as a cost-function we consider the specific rate~\cite{azuma2023quantum} defined as  
\begin{equation}\label{rate_formal}
    \mathcal{R} = f P_\text{succ}\; \frac{L/L_\text{att}}{T_\text{rep}mnN_\text{ph}},
\end{equation}
which also allows comparison with related studies~\cite{HANNES_PRX, ECONOMOU_PERFORMANCE}. In the equation above, $P_\text{succ}$ is the overall probability that the transmission of a single bit of information succeeds, and $f$ is the `secret key fraction'~\cite{ scarani2009security}, a figure of merit that captures the asymptotic performance of a specific QKD scheme over a given quantum channel; moreover, $T_\text{rep}$ is the duration of the repeater operation at each station, $n$ is the number of `matter qubits' employed per station, $N_\text{ph}$ is the number of photons per error-correcting code and $L_\text{att}\simeq 20\mathrm{km}$ is the typical attenuation length for telecom photons in optical fibers.\\
\indent We note that the division by $N_\text{ph}$ is not always present in literature~\cite{HANNES_PRX, ECONOMOU_PERFORMANCE} to avoid redundancy, as for many protocols one has $T_\text{rep}\simeq N_\text{ph}/\gamma$, where $\gamma$ is the single-photon emission rate~\cite{HANNES_PRX, ZHAN_PRL}; however, for reasons that will become clear later on, this is not the case in our scheme, where in fact one can even have $T_\text{rep} \gg N_\text{ph}/\gamma$, suggesting that the two figures of merit shall be highlighted separately in the optimization.  \\
\indent In this work, we improve on two key aspects that contribute to $\mathcal{R}$: (i) An efficient protocol for the hierarchical generation of LTCs; (ii) A novel optimization of the LTC structure. Connecting to the discussion above, the first is intended to improve $\mathscr{P}_\text{gen}$, while the second optimizes $\mathscr{E}_\text{QEC}$. In the following we provide a broad overview of the generation protocol and the optimization strategy, before entering more in detail in the sections below. 
\subsection{Hierarchical generation of loss-tolerant codes with a single quantum emitter}
Typical strategies for generating entanglement in photonic systems rely on optical setups where \emph{probabilistic} interactions between photons are engineered e.g. via non-linearities~\cite{milburn1989quantum}; in these settings, post-selection allows to implement universal quantum operations~\cite{knill2001scheme, bartolucci2023fusion}. Yet, the low success-rates constitute a key challenge in the near-term. In contrast, recent significant advances in the tunable coupling of controllable matter-based quantum systems with photonic degrees of freedom~\cite{uppu2021quantum, lodahl2015interfacing} are paving the way for a different perspective~\cite{javadi2015single}: experiments have demonstrated that entanglement between photons can be mediated \emph{deterministically} via the interaction of light and matter~\cite{le2022dynamical, reuer2022realization, ferreira2024deterministic}, also enabling the generation and manipulation of photonic entangled states \cite{Thomas_2024,ferreira2024deterministic, meng2023photonicfusionentangledresource,Huet_2025}. In these settings, the fundamental challenge becomes the matter-photon coupling.  \\
\indent For the generation of LTCs, we consider a paradigmatic quantum-optical setup, where one single emitter $\mathcal{Q}$ is coupled to a waveguide. Reducing to a single emitter is an important practical simplification: we note that many previous protocols for the generation of LTSs rely on several emitters~\cite{economou2010optically, gimeno2019deterministic}, or to one emitter coupled to several 
`memory' qubits~\cite{economou_prx, HANNES_PRX}. The emitter is driven by a time-dependent laser $\bm{E}(t)$, which constitutes the only control-knob on the system: all our proposed protocols are executed by simply modulating the phase- and time-profile of $\bm{E}(t)$. In addition, in our setting we consider a deterministic time-delayed quantum feedback~\cite{lloyd2000coherent} to the emitter, facilitated by a delay-line of fixed length: any emitted photon is fed-back to the emitter after a fixed time $\tau$, before entering the transmission line. Thus, our setup can be understood as depicted in Fig.~\ref{fig_1}(b): we control $\mathcal{Q}$ via the laser driving $\bm{E}(t)$, and at the output we collect the final desired entangled state in the photon field. Importantly, we do not assume any direct manipulation of the photons, nor measurements or resets of neither the photons nor the emitter.\\
\indent Our most important contribution on the generation of LTCs is a method for \emph{top-to-bottom} (i.e., hierarchical) generation. This can be understood by schematising the LTC with a tree-like graph as in Fig~\ref{fig_1}(a), where vertices represent photons and bonds represent entangling operations; this is the paradigmatic class of LTCs~\cite{VARNAVA_LOSS}. Loss-correction protocols proceed by performing \emph{adaptive} measurements, starting from the top vertex and proceeding towards the bottom~\cite{VARNAVA_LOSS, HANNES_PRX}. A crucial aspect to keep in mind is the following: since the loss probability is (exponentially) proportional to the raw time between the emission and the measurement of a photon, the performance of the loss-correction protocol is affected by the \emph{order} in which the photons are emitted~\cite{ECONOMOU_PERFORMANCE}. If they are emitted starting from the bottom layer of the tree-graph, the physical loss probability $p_\text{loss}$ to be corrected is enhanced to an effective rate
\begin{equation}\label{effective_p_loss}
    p'_\text{loss} \simeq 1-e^{-\gamma T_\text{LTC}} (1-p_\text{loss}),
\end{equation}
where $T_\text{LTC}$ is the overall time needed for generating the LTC and $\gamma$ is a loss rate. Previous proposals for deterministically generating LTCs are affected by this issue~\cite{economou_prx,HANNES_PRX,ZHAN_PRL}. Importantly, dedicated studies have found that the impact of the loss-enhancement above is significant~\cite{ECONOMOU_PERFORMANCE}, affecting specially the number of photons needed for correcting the losses, the required coherence of the emitters, and the communication rate. Although arbitrary photonic orderings can in principle be achieved with existing protocols~\cite{li2022photonic}, their application to LTCs would require a prohibitive overhead in the number of ancillary qubits. In contrast, our approach only makes use of one single emitter qubit. Finally, we note that the protocol in Ref.~\cite{ZHAN_PRL}, which is similar to some constructions presented below, could not natively allow for efficient hierarchical generation, as the native operation (an emitter-photon gate based on coherent feedback) only allows for the creation of one matter-photon entanglement bond per step. In contrast, below we will elaborate on methods that allow the creation of two all-photonic entanglement bonds per step; as we will show, this is sufficient to natively implement hierarchical generation.     \\
\indent Here, we harness a series of novel methods for designing an efficient protocol, that achieves \emph{top-to-bottom} generation in the paradigm described above. We then analyze the performance in loss-correction and long-distance quantum communication, assuming common quantum repeater schemes~\cite{HANNES_PRX}; our results are summarized in Fig.~\ref{fig_1}(c), where we report several improvements with respect to current approaches (note that, in the figure, the improvements due to top-to-bottom generation are combined with those discussed in the following paragraph). We broadly present these results in a later section.  
\subsection{Optimization of loss-tolerant codes via asymmetric graphs}
\indent Coming back to the analogy between LTCs and tree-graphs, we find it useful to divide the underlying tree in `principal branches' $k=0,1,\dots$ (counting from left) as shown in Fig.~\ref{fig_1}(a). We now recall the following important feature of loss-correction procedures. At the repeater stations, the recovery protocol~\cite{VARNAVA_LOSS, HANNES_PRX} prescribes that one considers the branches hierarchically, starting from the left; to each branch, we apply one among two possible procedures, dubbed by $\mathcal{X}$ and $\mathcal{Z}$, which we will summarize later on. Whether procedure $\mathcal{X}$ or $\mathcal{Z}$ is applied to branch $k$, it depends on the losses detected in the preceding branches $k'=0,1,\dots,k-1$. Crucially, the probability that we apply $\mathcal{X}$ to branch $k$ reads $p_\mathcal{X}(k) = (1-\epsilon)\epsilon^k$, where $\epsilon$ is the loss rate, while naively $p_\mathcal{Z}(k)=1-p_\mathcal{X}(k)$; that is, the probability of applying $\mathcal{X}$ is exponentially suppressed for the rightmost branches. This introduces a strong asymmetry in the tree-graph, as different branches will play (on average) drastically different roles during the recovery; noteworthy, typically the geometry of one branch can be optimized for \emph{either} the $\mathcal{X}$ or the $\mathcal{Z}$ operation, but \emph{not} for both at the same time. However, so far only symmetric tree-graphs were considered in the literature~\cite{VARNAVA_LOSS, HANNES_PRX, ECONOMOU_PERFORMANCE}.\\
\indent In contrast, here we take advantage of these asymmetries in the recovery protocol to design optimised underlying graphs for LTCs; specifically, by carefully selecting the geometry of each branch as function of $k$, we demonstrate higher recovery probabilities and repeater rates. This asymmetric design benefits the rate of secret bit transmission of Eq.~\eqref{rate_formal} in multiple ways. Most importantly, high-performance loss-correction can be achieved with fewer photons per LTC, resulting in faster generation times, further boosting the transmission efficiency. Consequently, the improved recovery probability also allows for greater spacing between repeater stations, extending the reach of communication. In the next sections, we provide an in-depth theoretical understanding of why asymmetric trees yield higher communication performance. We also show how this can lead to considerable improvements in the rate as a function of length, compared to protocols that utilize symmetric designs.

\subsection{Performance results: overview}
Before entering the details of the hierarchical generation and optimization in the sections below, we now anticipate a first comparison of our results with current state-of-the-art methods. Specifically, Fig.\ref{fig_1} (d) shows the behavior of the rate $\mathcal{R}$ for a fixed transmission line spanning $L= 300\; \mathrm{km}$, as a function of the number $m$ of repeater stations employed. The pink curve displays our results, while the blue and yellow ones are obtained via the generation method of Ref.~\cite{HANNES_PRX}; here we specifically consider LTCs featuring a maximum number of photons $N_\text{ph}\leq 150$~\footnote{In terms of LTCs as considered in Refs., the LTC maximizing $N_\text{ph}$ corresponds to a tree-graph with branching vector $\bm{b}=\left[5,5,5\right]$ - i.e., featuring $K=4$ layers with constant branching parameter $b_k=5$.} for the blue curve, and we do not set any maximum value $N_\text{ph}$~\footnote{Here we take into account all possible trees of depth 3, without imposing a maximum value for the branching parameter $b_k$} for the yellow one. We restrict to $m\leq 800$ signal repetitions (thus placing repeater stations not closer than $375\;\mathrm{m}$). From the numerical analysis, we see that in this setting our combined protocols achieve considerably higher communication rates, up to the regime of $\mathcal{R} \simeq 0.1 \mathrm{kHz}$ secret bits per repeater and employed photons~\footnote{To compare this with e.g. Ref.~\cite{HANNES_PRX}, it is important to recall that the rate considered here gives an explicit penalty on $N_\text{ph}$, the number of employed photons per LTC.}. These numerical results are obtained for an emitter operating at rates in the range $\gamma\sim 1 \mathrm{GHz}$, corresponding to a physical setup featuring e.g. a Silicon-vacancy color center coupled with a crystal cavity - a setup that has already been demonstrated in pioneering quantum repeater experiments~\cite{bhaskar2020experimental}. In a dedicated section we also explore alternative regimes and setups. Moreover, we will also discuss the robustness against finite coherence times of the emitter, which unavoidably reduces the fidelity of the LTC.

\section{Loss-tolerant codes: definitions}\label{section3}

Loss-tolerant codes can be understood as quantum \emph{graph codes} - a class of stabilizer codes with an immediate interpretation in terms of graphs~\cite{schlingemann2001quantum}. To introduce them, let $G=(V,E)$ be a graph, i.e. a set of vertices $v\in V$ and a set of edges $E$ between them. Then, the \emph{graph state} $\ket{\Phi_G}$ on $G$ is (constructively) defined as follows: (i) to each vertex $v\in V$ we assign a qubit $q_v$; (ii) we initialize each qubit in the state $\ket{+}=\left[\ket{0}+\ket{1}\right]/\sqrt{2}$; (iii) we apply a $\rr{CZ}_{i,j}=\mathbb{1}-2\ket{1_i1_j}\bra{1_i1_j}$ gate on each pair of vertices connected by an edge. Formally,
\begin{equation}\label{graph_state_definition}
    \ket{\Phi_G}=\left[\prod_{(i,j)\in E}\rr{CZ}_{i,j} \right] \bigotimes_{v\in V}\ket{+}_q.
\end{equation}
Graph states are thus manifestly stabilizer states, similarly to common quantum error-correcting codes; indeed, an alternative definition can be given by specifying the stabilizers. Note that, to be more precise, we can define a group of graph states, which is generated from $\ket{\Phi_G}$ by applying local $\rr{Z}_v$ operators on the vertices; these have the very same entanglement properties of $\ket{\Phi_G}$, and are equivalent to our scope. For this reason, we will always ignore corrections consisting of the application of $\rr{Z}$ operators on the vertices~\footnote{In principle, the relevant properties of graph codes are invariant under the action of the whole Pauli group; however, in our physical setting computational basis states play an important role during the feedback, rendering our protocols susceptible to $\mathrm{X}$ modifications.}.\\
\indent A graph code based on the graph state $\ket{\Phi_G}$ formally encodes a logical qubit as follows. Let $q$ and $\ket{\psi}_q$ be the qubit and the state we want to encode, respectively. Let also $V_q\subset V$ be a subset of the vertices of the associated graph $G$. Then, the encoding proceeds by (i) applying a $\mathrm{CZ}_{q,v}$ gate between $q$ and each vertex in $V_q$, and subsequently (ii) measuring $q$ in the $\mathrm{X}$ basis. Essentially, 
\begin{equation}
   \ket{\psi|\Phi_G} = \frac{\mathbb{1}\pm\mathrm{X}_q}{2} \Big[\prod_{v\in V_e} \mathrm{CZ}_{v,q} \Big]\ket{\psi}_q\otimes\ket{\Phi_G}
\end{equation}
is a state with support on the graph $G$~\footnote{According to our definition, $\ket{\psi|\Phi_G}$ actually has also support on $q$; however due to the measurement $q$ is in a disentangled state, which is known from the measurement outcome.}, containing the whole information about the original state $\ket{\psi}$; however, the information is now spread over the graph, and stored non locally in the correlations between the vertices. Then, the information can be retrieved by adequately measuring the graph. Crucially, for certain appropriate choices of $G$, the information $\ket{\psi}$ can be retrieved even in presence of losses in the graph; specifically, even if a subset $V_\text{lost}\subset V$ of the vertices is lost after the encoding, still one can restore the original state $\ket{\psi}$, given that $V_\text{lost}$ is not too large~\cite{VARNAVA_LOSS, bell2023optimizing}. This requires to perform measurements, which progressively reveal the locations of the losses, and to adapt the measurement bases with respect to these locations: in this sense, such protocols for loss correction are often called \emph{counterfactual}. Graph codes with this property are a suitable choice for a LTC.

\subsection{Loss-tolerance via tree-graph codes}
The paradigmatic example of LTC is given when the underlying grap $G$ is a \emph{tree-graph}. These LTCs were proposed in the seminal work by Varnava and colleagues~\cite{VARNAVA_LOSS}, where the loop-free structure of tree-graphs was first shown to be suitable for counterfactual loss-correction. After that, tree-graph codes have been used e.g as a fundamental underlying component for the design of all-optical (two-way) quantum repeaters~\cite{azuma2015all}, and as the key building block for one-way repeaters (where the tree-graph is the true main protagonist of the repeater scheme)~\cite{HANNES_PRX}.  \\
\indent We now introduce a convenient notation for describing the tree-graphs of interest. Specifically, here we consider tree-graphs composed by a \emph{root} vertex, $v_0$, plus a set of $K$ subsequent layers; we label the layers by $k\in\left\{0,1,...,K\right\}$, where layer $k=0$ contains only the root. For the encoding, we set $V_q\equiv\left\{v_0\right\}$. We refer to $K$ as the \emph{depth} of the tree-graph throughout the paper. Moreover, a \emph{symmetric} tree-graph is fully characterized by a branching vector $\bm{b}=\left\{b_0,b_1,...,b_{K-1}\right\}$, with $b_K=0$ by definition; $b_k$ is the number of `sons' attached to each vertex of layer $k$. We denote by $B_k$ the number of vertices of layer $k$, which thus satisfies $B_k=b_{k-1}B_{k-1}$, with $B_0=1$. We can address each vertex $v$ of the tree-graph by three coordinates, $v=(k,p,a)$: therein, $k$ identifies the level of $v$, $p\in\left\{ 0,...,B_{k-1}-1 \right\}$ identifies the position of its parent inside level $k-1$, and $a\in\left\{0,...,b_{k-1}-1\right\}$ the position of $v$ among all the children of $p$ (counting from left to right). We set $(0,0,0)$ to be the root by definition. It is useful to note that, with this notation, the position of $(k,p,a)$ inside layer $k$ is given by $b_{k-1}p+a$, meaning that its children $(k+1,p',a')$ will have
\begin{equation}
    p' = b_{k-1}p+a,
\end{equation}
with $a'$ being the free parameter. In the following, we will be particularly interested in binary-tree LTCs; these trees are such that $b_k=2$ for each layer, except the last one, $b_K=0$. Then, one has $B_k=2^k$, and the total number of vertices is $2^{K+1}-1$.\\
\indent The formalism above is well suited for addressing the vertices of any tree-graph; however, in the case of \emph{asymmetric} graphs, the tree-graph can no longer be characterized by a branching vector $\bm{b}$, and an exponential number of parameters are necessary in the most general case. We come back to this in a dedicated section, where asymmetric graphs play an important role. 

\section{Hierarchical generation of loss-tolerant codes}\label{section4}

We now enter the details of our first set of results, which concern the hierarchical \emph{generation} of LTCs. This section is divided in parts as follows. We start with an introduction to the quantum-optical setup considered, giving an overview on the main features and on the basic operations that can be implemented; then, we present a novel series of \emph{parallel entangling gates} based on the built-in feedback mechanism of our physical setting; finally, we leverage this to develop our hierarchical generation protocols. 

\subsection{Physical setup}
The central ingredient of our setting is a quantum emitter $\cl{Q}$ coupled to a waveguide; moreover, due to a feedback mechanism any emitted photon interacts again with $\cl{Q}$ once after the emission, with a time delay $\tau$. In all our protocols, we control directly only $\mathcal{Q}$; photons are never directly manipulated. In the next paragraph, we give a detailed account of the main features of the setup. Specifically, we first introduce the physical system, and then we discuss how the laser-driving $\bm{E}(t)$ can be employed for emitting photonic qubits; finally, we review how photons can deterministically interact with $\mathcal{Q}$. We postpone to Section~\ref{hardware considerations} the discussion of possible physical implementations.    

\subsubsection{Quantum emitter and photonic qubits}
We consider an emitter $\cl{Q}$ with the level structure shown in Fig.~\ref{fig_1}(a). Two metastable states, $\ket{g_0}$ and $\ket{g_1}$, compose a matter qubit; by tuning the field $\bm{E}(t)$ close to resonance to this transition, we can drive Rabi oscillations between them with Rabi frequency $\Omega_{01}(t)$. In addition, $\ket{g_1}$ can be coupled  with a third state $\ket{g_2}$ with frequency $\Omega_{12}(t)$; in turn, $\ket{g_2}$ can be excited to an excited state $\ket{e}$. The latter decays to $\ket{g_1}$ with rate $\gamma$; we can drive the excitation $\ket{g_2}\rightarrow\ket{e}$ with frequency $\Omega_{2e}(t)$. We define the Pauli algebra of the matter qubit with the convention $\rr{Z}=\ket{g_0}\bra{g_0}-\ket{g_1}\bra{g_1}$, $\rr{X}=\ket{g_1}\bra{g_0}+\ket{g_0}\bra{g_1}$.\\
\indent We define logical photonic qubits via time-bin encoding, as in Fig.~\ref{fig_sequences}(c): to each qubit $q$ we assign two time slots (early and late) and define the logical states accordingly to the presence of one photon in the early ($\ket{1}$ state) or late bin ($\ket{0}$ state). Note that a photon must \emph{always} be present in one of the two bins - thus the absence of the photon (e.g., during a measurement) heralds the loss of the qubit. Graphically, we represent a time-bin qubit as in Fig.~\ref{fig_sequences}(c), bottom line. 
\subsubsection{Emission sequences}
\begin{figure}[t!]
    \centering
    \includegraphics[width=1.0\linewidth]{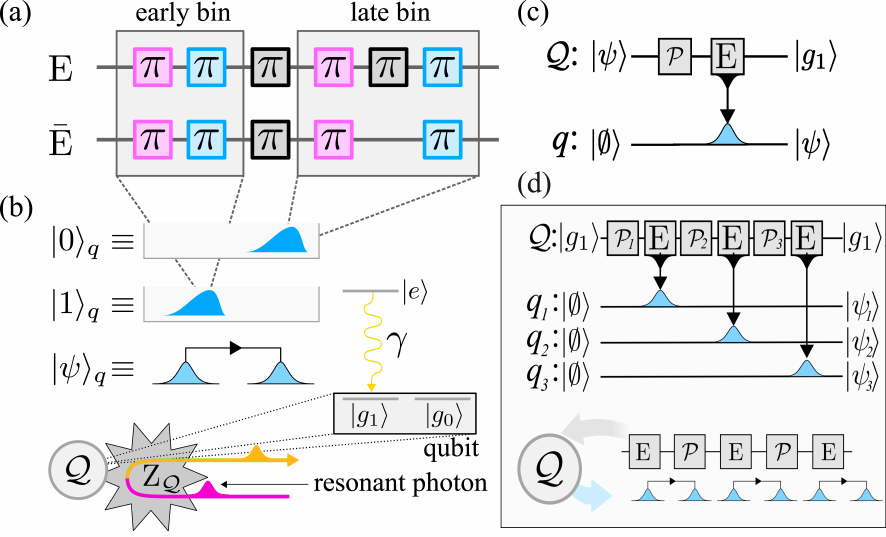}
    \caption{(a) The emission sequences which enable to prepare logical photonic qubits via the emitter: $\mathrm{E}$ entirely transmits the quantum state from $\mathcal{Q}$ to the photonic degree of freedom, while $\Bar{\mathrm{E}}$ generates a light-matter entangled state. (b) The definition of the logical photonic time-bin qubit; below, we show how we represent graphically a generic qubit. (c) A fed-back photon on resonance with the $\ket{g_1}\leftrightarrow\ket{e}$ transition couples with $\mathcal{Q}$ if the latter is in $\ket{g_1}$; thus the conditional scattering phase triggers a logical $\mathrm{Z}_\mathcal{Q}$ gate. } 
    \label{fig_sequences}
\end{figure}

Photonic qubits can be prepared in arbitrary states by manipulating the emitter; the main scheme is composed by (i) a preparation operation $\mathcal{P}$, which initializes $\cl{Q}$ in a desired state~\footnote{In our case, the $\cl{P}$ operation will always be a gate. Specifically, in our protocols we will always know the initial state of $\cl{Q}$ before the emission, so we just apply a unitary in order to rotate it to the target state. Alternatively, if the state is unknown, $\cl{P}$ must be a projection, which is however much slower to execute.}, and (ii) a pulse sequence on $\cl{Q}$ which transfers the information to the photonic degrees of freedom. We now introduce two such sequences, $\rr{E}$ and $\bar{\rr{E}}$, which are displayed in Fig.~\ref{fig_sequences}(a): the former completely transfers the state to a photonic qubit while resetting $\cl{Q}$; the latter keeps the two entangled.\\
\indent First, let us consider $\rr{E}$. Let the emitter be in an arbitrary qubit state $\ket{\psi}=\alpha\ket{g_0}+\beta\ket{g_1}$; we can understand the effect of $\mathrm{E}$ by linearity. If $\cl{Q}$ is initially in $\ket{g_1}$, the first pair of $\pi$ pulses excites it as $\ket{g_1}\rightarrow\ket{g_2}\rightarrow\ket{e}$, with subsequent decay back to $\ket{g_1}$ and emission of a photon in the \textit{early} time-bin (i.e., the $\ket{1}$ state); the subsequent $\pi$ pulse on the $\ket{g_0}\leftrightarrow\ket{g_1}$ coupling then flips the atomic populations, bringing $\cl{Q}$ in $\ket{g_0}$. The emitter is therefore unaffected by the subsequent pulse on the $\ket{g_1}\rightarrow\ket{g_2}$ coupling; then, the $\Omega_{01}(t)$ pulse brings it in $\ket{g_1}$, where it is again unaffected by the last driving. The only overall effect is therefore the emission of a photon in the early time-bin; vice versa, if $\cl{Q}$ is initially in $\ket{g_0}$, with similar arguments one can see that a photon is emitted in the \emph{late} time-bin ($\ket{0}$ state), but $\cl{Q}$ again ends up in $\ket{g_1}$. Altogether, by linearity any emitter's state $\ket{\psi}$ is entirely transferred to the photonic degree of freedom, with $\cl{Q}$ deterministically reset to $\ket{g_1}$:
\begin{equation}\label{machine_gun}
    \rr{E}\left[\ket{\psi}_\cl{Q}\ket{\emptyset}_q \right] = \ket{g_1}_\cl{Q} \ket{\psi}_q = \rr{SWAP} \!\left[ \ket{\psi}_\cl{Q}\ket{1}_q \right] ;
\end{equation}
therein, by $\ket{\emptyset}_q$ we denote the photonic vacuum, and we interpret the operation as a $\rr{SWAP}$ gate with the photonic qubit initialized in $\ket{1}$. In the following, we will represent this type of emission in a circuit form as in Fig.~\ref{fig_sequences}(b).\\ 
\indent We can therefore emit any (factorized) state $\ket{\psi_1\psi_2...\psi_N}$ with the following protocol: starting with $\cl{Q}$ in $\ket{g_1}$, we apply the operation $\rr{E}\cl{P}_N...\rr{E}\cl{P}_2\rr{E}\cl{P}_1$, where $\cl{P}_k$ maps $\ket{g_1}_\cl{Q}\rightarrow\ket{\psi_k}_\cl{Q}$. Formally, we have
\begin{equation}    \ket{g_1}_\cl{Q}\bigotimes_{k=1}^N\ket{\psi_k}_{q_k} = \rr{E}\cl{P}_N...\rr{E}\cl{P}_1\ket{g_1}_\cl{Q}\bigotimes_{k=1}^{N}\ket{\emptyset}_{q_k},
\end{equation}
where $q_1,...,q_N$ label the photonic logical qubits in consideration. This effectively realizes a so-called \emph{quantum machine gun} for product states; we represent it as shown in Fig.~\ref{fig_sequences}(d).\\
\indent Second, we consider the $\bar{\rr{E}}$ sequence. The difference with $\rr{E}$ is that we avoid the second $\pi$ pulse on the $\ket{g_0}\leftrightarrow\ket{g_1}$ transition before the last excitation pulse. Thus, $\cl{Q}$ is not reset at the end of the sequence, but remains entangled with the photonic qubit; specifically, any emitter's state $\ket{\psi}=\alpha\ket{g_0}+\beta\ket{g_1}$ is mapped to the entangled state $\alpha\ket{g_1}\ket{0}+\ket{g_0}\ket{1}$. This can be written as 
\begin{equation}
    \bar{\rr{E}}\left[\ket{\psi}_\cl{Q}\ket{\emptyset}_q\right] =  \rr{X}_\cl{Q} \circ\rr{CX} \left[\ket{\psi}_\cl{Q}\ket{0}_q\right],
\end{equation}
which is locally equivalent to a $\rr{CX}=\ket{g_0}\bra{g_0}+\ket{g_1}\bra{g_1}\otimes\rr{X}_q$ gate with the photonic qubit initialized in $\ket{0}_q$. Adding the sequence $\bar{\rr{E}}$ to the machine gun enlarges the family of achievable states to a wide class of entangled states with 1D entanglement connectivity; this includes, for instance, GHZ states and linear cluster states~\cite{lindner2009proposal}. We will represent emissions through $\bar{\mathrm{E}}$ analogously as for Fig.~\ref{fig_sequences}(b), by simply writing $\bar{\mathrm{E}}$ in place of ~$\mathrm{E}$.\\
\subsubsection{Feedback mechanism}
In our architecture, we realize a time-delayed feedback of the emitted photons to $\cl{Q}$; specifically, each photon, after the emission, enters a delay line and then interacts with $\cl{Q}$ after a fixed time delay $\tau$, before being injected in the main waveguide. The delay $\tau$ is therefore a fixed parameter in our setup. The trajectory of a photon after the emission is sketched in the space-time diagram in Fig.~\ref{fig_1}(b); we use different colors (blue, pink, orange) for representing the three stages a photon goes through: respectively, after the emission, before the feedback, after the feedback. We remark that (i) each photon is fed-back exactly once, (ii) the delay $\tau$ of the feedback is fixed for every photon, and (iii) no operation is executed directly on the photon except the feedback. Graphically, we insert a component $\mathcal{M}$ in the waveguide in Fig.~\ref{fig_1}(c) for representing the feedback operation.\\
\indent Under the lens of fundamental physics, this paradigm has been studied extensively, highlighting nontrivial effects arising due to the non-Markovian character of the resulting dynamics of the radiation field~\cite{grimsmo2015time, ask2022non, vodenkova2024continuous}. Crucially, this non-Markovianity in fact equips a simple system with a natural memory, that can be leveraged to prepare nontrivial 2D entangled states~\cite{pichler2017universal, shi2021deterministic}, including bottom-to-top LTCs~\cite{ZHAN_PRL}. We finally remark that this setup has recently been demonstrated in the microwave regime, where an additional qubit was controlled to act as a switchable mirror~\cite{ferreira2024deterministic}.  \\

\subsubsection{Emitter-photon interactions} 
When a photon is fed-back, interactions allow for engineering entanglement between photonic qubits and $\mathcal{Q}$. To understand this, let us consider the situation displayed in Fig.~\ref{fig_sequences}(e): an incident photon is fed-back to $\cl{Q}$, which can be in any qubit state $\ket{\psi}=\alpha\ket{g_0}+\beta\ket{g_1}$. In the following, we consider unidirectional coupling and thus neglect mode losses during the coupling (i.e., we assume that the emission rate $\gamma$ in the guided mode is larger than the emission rate $\gamma_\text{else}$ in any other mode). We assume the photon is resonant with the $\ket{g_1}\leftrightarrow\ket{e}$ transition, and does not couple to $\ket{g_0}$; this occurs e.g. if it was emitted via the decay $\ket{e}\rightarrow\ket{g_1}$ as in the previous protocols. Then, the photon interacts with $\cl{Q}$ only if the latter is in $\ket{g_1}$. Remarkably, \emph{if} the photon wave-packet has narrow bandwidth $\cl{B}$ with respect to the coupling strength, i.e. $\cl{B}\ll\gamma$, then the system picks up a $-1$ phase due to the interaction~\cite{shen2005coherent, ralph2015photon}; this occurs if and only if $\cl{Q}$ is in $\ket{g_1}$, effectively resulting in the following unitary gate applied to the emitter:
\begin{equation}
    \mathrm{Z}_\mathcal{Q} = \mathbb{1}-2\ket{g_1}\bra{g_1}. 
\end{equation}
With the assumption that $\cl{Q}$ is initially in the qubit subspace $\left\{\ket{g_0},\ket{g_1}\right\}$, this effectively implements a `logical' $\mathrm{Z}$ gate to the qubit. By exploiting this phenomenon, in the following we will engineer effective all-photonic entanglement by subssequently transferring the state to the photonic degrees of freedom.\\
\indent Importantly, the deterministic $-1$ phase is only triggered when the condition $\cl{B}\ll\gamma$ is met; otherwise, the wave-packet is distorted during the scattering. However, a photon naively emitted during the $\ket{e}\rightarrow\ket{g_1}$ decay does \emph{not} satisfy this requirement, as the wave-packet would be Lorentzian with bandwidth $\mathcal{B}\sim \gamma$. In order to realize narrow bandwidth wave-packets, when required we drive the excitation $\ket{g_2}\rightarrow\ket{e}$ by turning on and off $\Omega_{2e}(t)$ \emph{adiabatically}; this allows to shape the temporal profile of the emitted photon, and particularly to control its bandwidth $\cl{B}$~\cite{pichler2017universal}. \\
\indent Due to adiabaticity, the duration of the sequence becomes longer; it is thus important to distinguish the cases in which this slower protocol is employed: we label by $\rr{S}$ and $\bar{\rr{S}}$, respectively, the `slow' versions of $\rr{E}$ and $\bar{\rr{E}}$. All the graphical notations, such as in Fig.~\ref{fig_sequences}(b), are straightforwardly adapted for these emissions, by simply writing the correct symbol in place of $\mathrm{E}$. As a figure of merit, we consider the ratio between the duration of a slow emission, $\tau_\rr{S}$, and the duration of a fast emission, $\tau_\rr{E}\simeq 1/\gamma$, and quantify it by the factor $\beta = \tau_\rr{S}/\tau_\rr{E}$; for the fidelity of the gates to be high (i.e., above the $99\%$ range), this factor must be of order $\beta \simeq 500$~\cite{ECONOMOU_PERFORMANCE}: thus, the amount of slow emissions employed drastically affects the rates of the protocols, and must be taken in account.

\subsection{Parallel entangling gates}

\begin{figure}[t!]
    \centering
    \includegraphics[width=1.0\linewidth]{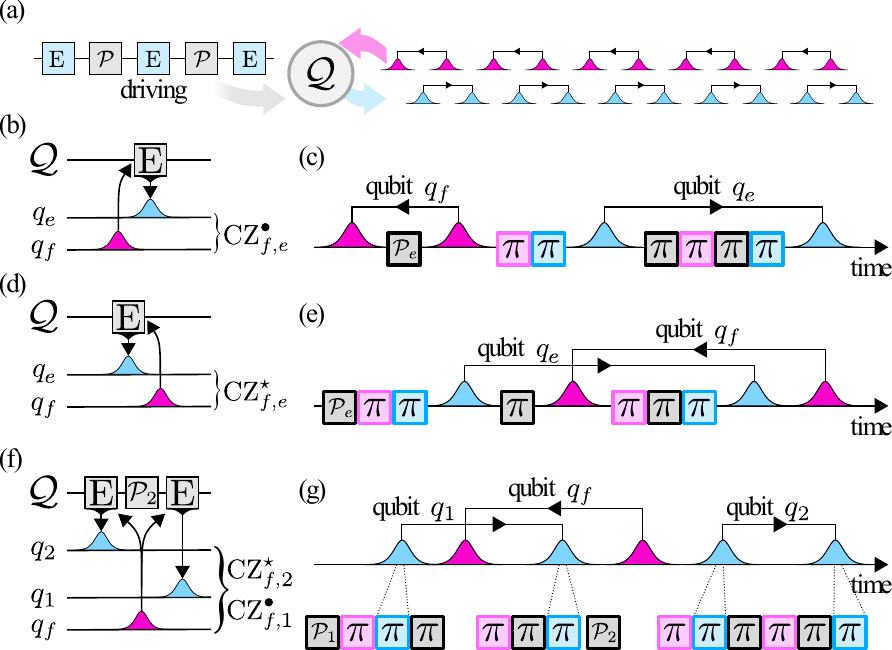}
    \caption{(a) While the driving triggers the emitter $\cl{Q}$ to emit photonic qubits (in blue), previously emitted qubits are fed-back to it (in red). (b, d) Circuit representation of the entangling gates $\mathrm{CZ}^\bullet$ and $\mathrm{CZ}^\star$. (c,e) The relative timings between the emitted qubit $q_e$ and the fed-back one $q_f$ which enable the gates. (f) The two gates can be merged together in parallel, entangling one fed-back qubit $q_f$ with two emitted qubits $q_1$ and $q_2$. (g) The explicit timing which enables the gates to work in parallel.   }
    \label{fig_gates}
\end{figure}

We now move to the design of parallel entangling gates: specifically, we engineer the feedback in such a way that all-photonic logical gates emerge naturally. We are interested in the main situation displayed in Fig.~\ref{fig_gates}(a): while the machine-gun driving
\begin{equation}
...\circ\mathcal{P}\circ\mathrm{E}\circ\mathcal{P}\circ\mathrm{E}\circ...,
\end{equation}
is applied to $\mathcal{Q}$, which thus keeps on emitting photonic qubits (in blue), a string of previously emitted qubits (in red) is fed-back to it, interacting with $\mathcal{Q}$ during the emission process. We show below that, depending on the relative \emph{timing} between the emission of one qubit and the feedback events, effective entangling gates emerge; in particular, we identify two relative timings, between the emission sequence of a qubit $q_e$ and the feedback of a previously emitted qubit $q_f$, which give rise to $\mathrm{CZ}-$like gates between the two. For convenience, we will refer to these crucial timings as `methods' (for entanglement generation), and we label them as $\mathrm{CZ}^\star_{f,e}$ and $\mathrm{CZ}^\bullet_{f,e}$. The two methods are \emph{only} defined by the relative timings, and do not require additional operations during the driving; importantly they do not exclude each other. Crucially, during the feedback of $q_f$, we can use \emph{both} methods simultaneously for entangling $q_f$ in parallel with two sequentially emitted photonic qubits, $q_1$ and $q_2$, thus realizing a gate of the form $\mathrm{CZ}^\bullet_{f,1}\mathrm{CZ}^\star_{f,2}$. Below, we explain these methods in detail, and how they can be combined.  \\

\subsubsection{First method for entanglement generation}
In order to illustrate the idea for the first method, consider the situation displayed in Fig.~\ref{fig_gates}(c), where we represen the events occurring at $\mathcal{Q}$ in time: here the feedback of a time-bin qubit $q_f$ is nested during the sequence $\mathrm{E}\circ\mathcal{P}_e$, designed to emit a qubit $q_e$~\footnote{Note that $q_e$ might be emitted with any of the sequences presented above, so we consider $\rr{E}$ for concreteness.}. We assume the condition $\cl{B}\ll\gamma$ is met for $q_f$; thus, the resonant coupling of $q_f$ with the $\ket{g_1}\leftrightarrow\ket{e}$ transition triggers conditional phases, which have different logical impacts depending on the timing. Specifically, for the first method we choose the following timing [see Fig.~\ref{fig_gates}(c)]: (i) $\cl{P}_e$ is applied in between the feedback of the two time-bins of $q_f$ (i.e., after the early bin has interacted with $\cl{Q}$, but before the late one has), and (ii) $\rr{E}$ is executed only after the feedback of the late bin of $q_f$. This way, $\cl{Q}$ is in the state $\ket{\psi_e}$ only during the interaction with the late bin of $q_f$. Then, we have a coupling to the $\ket{g_1}\leftrightarrow\ket{e}$ transition if and only if both the following conditions are realized: (i) a photon is present in the late bin of $q_f$ (i.e., the logical qubit is in $\ket{0}_f$) \textit{and} (ii) the emitter is in $\ket{g_1}$ during the feedback of the late bin of $q_f$, which in turn happens if and only if $q_e$ is emitted in $\ket{1}_e$. Since the scattering provides a $-1$ phase, the overall operation can be written as an effective all-photonic two-qubit gate between $q_f$ and $q_e$:
\begin{equation}
    \rr{CZ^\bullet}_{f,e} = \ket{1}_f\bra{1} + \ket{0}_f\bra{0}\otimes \rr{Z}_e = \rr{Z}_e\rr{CZ}_{f,e},
\end{equation}
which is a $\rr{CZ}$ up to a local $\rr{Z}$ byproduct on $q_e$. Thus, this reciprocal timing between the feedback of $q_f$ and the emission of $q_e$ provides a first method to effectively entangle the two. In fact, the emitter $\cl{Q}$, during the emission process, \emph{mediates} an entangling gate between photonic qubits, thanks to the feedback; we represent this concept in a circuit form as in Fig.~\ref{fig_gates}(b). \\ 

\subsubsection{Second method for entanglement generation}
For the second method, we consider the same situation, but now with a different timing [see Fig.~\ref{fig_gates}(e)]: (i) $\cl{P}_e$ is applied, and $\rr{E}$ begins, before the feedback of $q_f$; (ii) the feedback is such that the early bin of $q_f$ interacts with $\cl{Q}$ before the emission of the late bin of $q_e$, but after the emission of the early one. This way, the $-1$ phase is provided if and only if $q_f$ is in $\ket{1}_f$ and $q_e$ is emitted in $\ket{0}_e$, leading to another phase gate,
\begin{equation}
    \rr{CZ}^\star_{f,e}= \ket{0}_f\bra{0} - \ket{1}_f\bra{1}\otimes \rr{Z}_e = \rr{Z}_f\rr{CZ}_{f,e},
\end{equation}
which is locally equivalent to the previous one. In summary, this other timing provides a second method to effectively entangle $q_f$ and $q_e$. Similarly, we represent it as a circuit as shown in Fig.~\ref{fig_gates}(d). \\

\subsubsection{Engineering parallel entangling gates} 
The crucial point is that the two methods above can be applied to the same fed-back qubit $q_f$ \emph{in parallel}. Specifically, let $q_1$ and $q_2$ be two qubits to be emitted sequentially in this order; then, we can set the timing of the feedback of $q_f$ in such a way that both methods apply, realizing respectively $\rr{CZ}^\star_{f,1}$ and $\rr{CZ}^\bullet_{f,2}$. This is achieved by the following timing (let $\ket{\psi_1}$ and $\ket{\psi_2}$ be the target states for $q_1$ and $q_2$ respectively); see Fig.~\ref{fig_gates}(g) for a detailed reference. (i) First, apply $\cl{P}_1$ and start $\rr{E}$; (ii) the early time bin of $q_f$ interacts with $\cl{Q}$ after the emission of the early bin of $q_1$; (iii) $\cl{P}_2$ is applied after the emission of the late bin of $q_1$; (iv) then the late bin of $q_f$ interacts with $\cl{Q}$; (v) finally, $\rr{E}$ is applied again to emit $q_2$. Altogether, the resulting gate is
\begin{equation} \label{methods_combined}   \rr{CZ^\bullet_{f,2}}\rr{CZ^\star_{f,1}} = \rr{Z}_2\rr{Z}_f \rr{CZ}_{f,1} \rr{CZ}_{f,2},
\end{equation}
effectively realizing two controlled-phase gates in parallel. As we explain in the next section, this parallel execution of entangling gates during the feedback is exactly the building block which was missing for unlocking top-to-bottom generation of LTCs. At the cricuit level, we merge the previous notations as shown in Fig.~\ref{fig_gates}(f) for representing the combination $\rr{CZ^\bullet_{f,2}}\rr{CZ^\star_{f,1}}$. \\
\indent Since the entangling gates emerge as a consequence of the feedback during an emission sequence, it is convenient to introduce the following notation, which captures the effects at the logical level. We will denote by $\rr{E}^\star_{f,e}$ the case in which the first timing of the feedback of $q_f$ occurs, while $q_e$ is emitted through $\rr{E}$; specifically, we write
\begin{equation}
    \rr{E}^\star_{f,e}\left[ \ket{\psi}_f\ket{\phi}_\cl{Q}\ket{\emptyset}_e \right] = \ket{g_1}_\cl{Q}\otimes  \rr{CZ}^\star_{f,e}\left[\ket{\psi}_f\ket{\phi}_e\right].
\end{equation}
Similarly, we define $\rr{E}^*_{f,e}$, $\Bar{\rr{E}}^{\star/\bullet}_{f,e}$, and the corresponding slow versions when $\mathrm{S}$ or $\Bar{\mathrm{S}}$ are employed.\\ 
\indent Note that entanglement is generated \textit{if and only if} the correct time-bin concatenations are realized: one can also design the timings in such a way that no entanglement is generated at all. A nontrivial example is the following: suppose $q_f$ is fed-back during the emission of $q_e$ in such a way that (i) $\cl{P}_e$ is executed before $q_f$ interacts with $\cl{Q}$, but (ii) the $\rr{E}$ sequence emitting the early bin of $q_e$ is executed after the first bin of $q_f$ has interacted with $\cl{Q}$. Then, one essentially obtains both entangling gates between $q_f$ and $q_e$, resulting in the local phases
\begin{equation} \label{methods_distructive}   \rr{CZ}^\bullet_{f,e}\rr{CZ}^\star_{f,e} = \rr{Z}_e\rr{Z}_f;
\end{equation}
effectively, when this other timing of the feedback is employed, the two entangling gates annihilate. Similarly, other timings can be engineered, such that effectively no interaction occurs. These observations, in addition to stressing the importance of the proposed time-bin concatenation, also provide an important degree of freedom while designing protocols for the parallel generation of large graph states; indeed, in specific circumstances one might need to \textit{avoid} generating determinate bonds during the feedback.\\
\indent Finally, we remark that the timing is only set by the control on $\cl{Q}$, i.e., by executing the pulse sequences at the correct times; specifically, given the times of the feedback of the two bins of $q_f$, one can always adapt the emission of $q_e$ in such a way that the desired reciprocal timing is realized. This allows to execute all the protocols described below with a \emph{static} architecture, were the time-delay of the feedback $\tau$ (i.e., the length of the delay line) is fixed, without the need of introducing further delays or operations on the photons: all the steps are executed by adequately driving $\cl{Q}$. Moreover, as discussed above, one can also set the reciprocal timings in such a way that no entanglement arises at all; thus, our protocols are compatible with a scenario in which \emph{all} photons are fed-back, thus not requiring to exclude some of them from the delay-line. 

\subsection{Pilot protocol for top-to-bottom generation}

We now employ the methods introduced above to design efficient protocols to generate LTCs; most importantly, these protocols are \emph{hierarchical}, i.e. they proceed in a \emph{top-to-bottom} fashion, thus following the logic of the recovery protocol. In the following, we outline a \emph{pilot protocol}, which illustrates our generation method on the simplest (yet significant) example; building on this, we discuss how the pilot protocol is modified to generate more generic LTCs. 

\begin{figure*}[t!]
    \begin{minipage}{1.0\textwidth}
        \includegraphics[width=\textwidth]{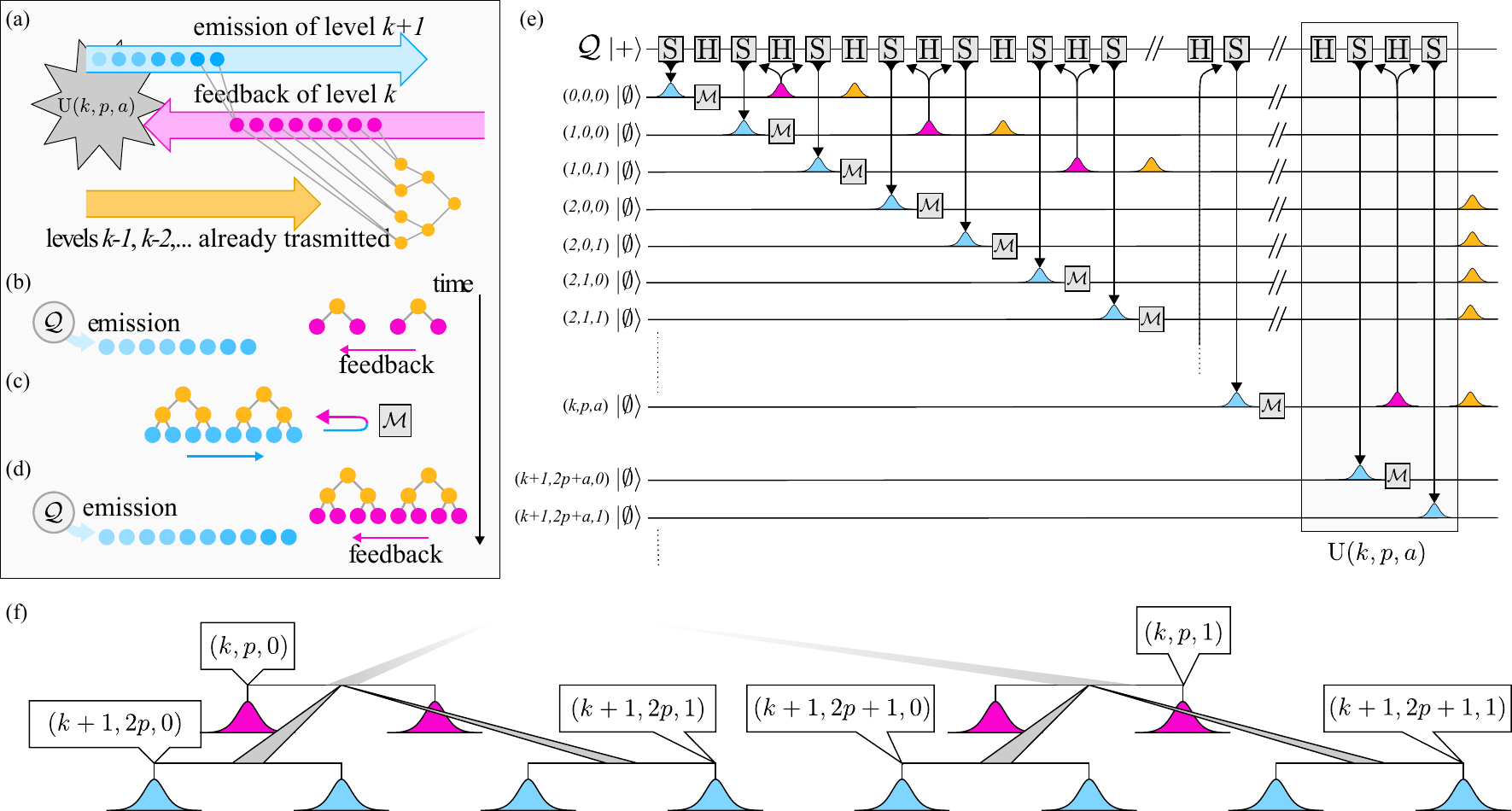}
    \end{minipage}%
\caption{(a) The main concept of the pilot protocol: one layer of the tree ($k+1$ in the picture) is emitted while the layer above ($k$) is fed-back to the emitter; the scattering happens simultaneously with the emission process, triggering parallel entangling gates, such that each fed-back qubit of level $k$ is entangled with two qubits of level $k+1$. (b-d) Details of the protocol. (b) One layer is emitted (in blue) while the previous one is fed-back (in red). (c) The fed-back level (now in green, as the feedback has already occurred) is entangled with the blue one; the latter hits the mirror $\mathcal{M}$, thus preparing for being fed-back (and becoming red). (d) Now the next layer is emitted, and the process restarts. (e) Circuit representation of the pilot protocol. (f) Detailed picture of the timings required for the tree-LTC to emerge; in the picture, the feedback of $(k,p,0)$ and $(k,p,1)$ are represented, triggering the unitaries $\mathrm{U}(p,k,0)$ and $\mathrm{U}(p,k,1)$. For a detailed labeling of the vertices according to the emission order, we direct to Fig.~\ref{tree_scheme} in Appendix~\ref{modification to the pilot}. }
\label{fig_pilot}
\end{figure*}

\subsubsection{Code generation: pilot protocol}
The pilot protocol straightforwardly generates LTCs whose underlying graph structure is a binary-tree; crucially, the emission proceeds naturally \emph{from top to bottom}, i.e. emitting the root $v_0=(0,0,0)$ of the tree-graph first, and then proceeding for increasing layers $k$ until the last layer $K$, which is emitted last. We can summarize it in the following points (also refer to Fig.~\ref{fig_pilot}). (i) We sequentially emit the layers of the tree, starting from the top layer (i.e., from the root); each qubit in each layer is fed-back exactly once after the same, fixed delay $\tau$. (ii) Each qubit is prepared in $\ket{+}$ through the emission sequence $\rr{S}$; thus, the machine-gun driving on the emitter takes the form
\begin{equation}\label{pilot_machine_gun}
    ...\mathrm{S}\circ\mathrm{H}\circ\mathrm{S}\circ\mathrm{H}\circ\mathrm{S}\circ\mathrm{H}...,
\end{equation}
with $\mathcal{Q}$ initialized in $\ket{g_1}$. 
(iii) While layer $k+1$ is being emitted, layer $k$ is fed-back, interacting with $\cl{Q}$ during the emission process; the timing is engineered by controlling the driving, i.e., by judiciously setting the emissions via the shape of $\bm{E}(t)$. (iv) Each qubit of layer $k$ is fed-back in such a way to become entangled with two consecutive qubits of layer $k+1$, by employing in parallel the two methods presented above. Specifically, consider a qubit $(k,p,a)$; then, we set the feedback in such a way that it gets entangled with its two children, $(k+1, 2p+a, 0)$ and $(k+1,2p+a,1)$, through $\rr{CZ}^\bullet$ and $\rr{CZ}^\star$ respectively, thereby realizing 
\begin{equation}\label{nested_CZ}
\mathrm{CZ}_{(k,p,a), (k+1,2p+a, 0)}^\bullet  \mathrm{CZ}_{(k,p,0), (k+1,2p+a,1)}^\star,  
\end{equation}
which is nested in the process~\eqref{pilot_machine_gun}. This is achieved by employing the overall timing displayed in Fig.~\ref{fig_gates}(f), which ensures the desired timing for each feedback.\\ 
\indent We can formally describe the pilot protocol as follows. First, for each qubit in layers $k\leq K-2$ we define 
\begin{align}
    \rr{U}(k,p,a)= &\; \rr{S}^\star_{(k,p,a),(k+1,2p+a,1)}\rr{H}_\cl{Q}\times \nonumber \\
    & \times \rr{S}^\bullet_{(k,p,a),(k+1,2p+a,0)}\rr{H}_\cl{Q} \;\;\;\text{for}\;\;\;k\leq K-2,
\end{align}
which describes the feedback of $(k,p,a)$ during the emission of its children. Essentially, $\mathrm{U}(k,p,a)$ captures the action of the gates~\eqref{nested_CZ} nested in the sequence~\eqref{pilot_machine_gun}. Differently, for layer $k=K-1$, we set
\begin{align}
    \rr{U}(k,p,a)= & \;\rr{E}^\star_{(k,p,a),(k+1,2p+a,1)}\rr{H}_\cl{Q}\times \nonumber \\
    & \times \rr{E}^\bullet_{(k,p,a),(k+1,2p+a,0)}\rr{H}_\cl{Q} \;\;\;\text{for}\;\;\;k = K-1;
\end{align}
this is because the last layer qubits do not need to interact again with $\cl{Q}$, so we can generate them through $\rr{E}$. The last layer, $K$, does not need to be fed-back, so we do not need to define a unitary $\mathrm{U}(K,p,a)$. Then, the overall operation is described by
\begin{equation}\label{pilot_unitary}
    \rr{U}_\text{pilot}=\prod_{k=0}^{K-1} \prod_{p=0}^{2^{k-1}-1} \prod_{a=0}^{1} \rr{U}(k,p,a);
\end{equation}
by initializing $\cl{Q}$ in $\ket{g_1}$, application of this leads to
\begin{equation}
    \rr{U}_\text{pilot} \ket{g_1}\otimes \ket{\emptyset}_\text{tree} = \ket{g_1}\otimes\ket{\Phi}_\text{tree},
\end{equation}
where the equality is intended up to local $\rr{Z}$ corrections, which we ignore, as mentioned above, since they do not change the properties of the state.\\
\indent In conclusion, the pilot protocol generates binary-tree LTCs from top to bottom in the proposed static setting. These are the simplest type of LTC; while they are not optimal in general, due to their simplicity they are mostly appealing for experimental realizations, specially in the near-term. 


\subsubsection{Generic generation protocols and optimizations: summary}

In addition to the simple example of binary-tree LTCs, our methods can be employed to generate a broader family of tree-graph LTCs, by suitably modifying the pilot protocol presented above. Specifically, several branching parameters can be adjusted to hierarchically construct a wide range of tree structures. The family of LTCs that can be generated in this way is not universal, yet it is sufficiently rich to allow for substantial improvements in practical applications—e.g., in quantum repeater protocols, as we will show in the following sections. More precisely, our top-to-bottom approach allows for generating trees with arbitrary depth, and arbitrary number of photons in the final layer. The main structural constrain lies in the middle layers of the trees, where the branching element is limited to $b_k\leq 4$ for the internal branches and to $b_k\leq 3$ for the external ones. Importantly this class of tree-graph LTCs gives us enough freedom to optimize various figures of merit, e.g., the generation-rate (i.e., the time needed for preparing a LTC). These modified methods are based on the very same concepts presented here, but require rather technical explanations; we therefore detail them extensively in Appendix~\ref{modification to the pilot}.


\section{Optimizing the loss-tolerance via asymmetric codes}\label{section5}

We now present the second key contribution of our work, which concerns the optimization of LTCs via asymmetric tree-graphs. In the following, by `asymmetric' tree-graph we mean the following concept. First, recall that a \emph{symmetric} tree-graph is fully-characterized by the branching vector $\bm{b}=\left(b_0,b_1,\dots,b_{K-1}\right)$, where $K+1$ is the depth of the tree-graph, and $b_k$ is the number of `son' vertices attached to each vertex of layer $k$. In contrast, an asymmetric tree-graph can be specified via a collection $\bm{B}=\left(\bm{B}_0,\bm{B}_1,\dots,\bm{B}_{b-1}\right)$. Herein, $b$ is the branching parameter of the root vertex $v_0$; the parameter $\bm{B}_k$ specifies the geometry of the branch $k$. Note that each of these branches has again the shape of a tree-graph; thus, the entire graph can be specified by iterating this formalism. We note that, clearly, in the most general case the number of parameters eventually becomes exponentially larger than in the symmetric case. Moreover, the tree-graph is symmetric if the branches are all equal: $\bm{B}_0=\bm{B}_1=\dots =\bm{B}_{b-1}$; in any other case, the graph is referred to as asymmetric.   \\ 

\begin{figure*}[t!]
    \begin{minipage}{1.0\textwidth}
        \includegraphics[width=\textwidth]{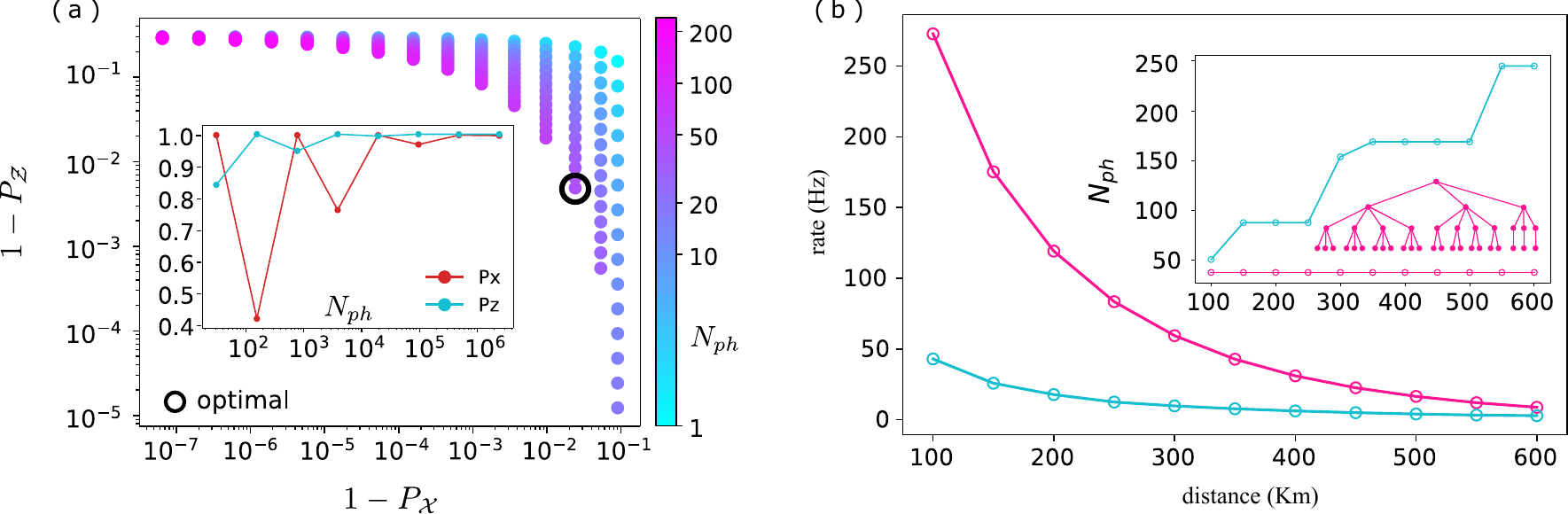}
    \end{minipage}%
\caption{(a) In the main plot, every point represents a possible branch of a tree-graph LTC, positioned according to the coordinates $\left(1- P_{\mathcal{X}},1- P_{\mathcal{Z}}\right)$. Here we consider tree-graphs with maximum branching number $15$ and depth $2$ (corresponding to LTCs of depth $3$). Although apparently small, we remark that these LTCs have been identified as optimal for several specific implementations in previous works~\cite{HANNES_PRX,ECONOMOU_PERFORMANCE}. The color indicates the number of photons composing the graph (i.e., of photons in the LTC branch). The circled point corresponds to  the optimal symmetric LTC, which minimizes $1-P_{rec}$ for $\epsilon = 0.3$. Inset: the success probability of $P_\mathcal{X}$ and $P_\mathcal{Z}$, with $\epsilon = 0.3$, for LTCs with constant branching vector (picked to be $5$ here), as a function of the depth of the related tree-graph. (b) In the main plot the maximized rate $\mathcal{R}$, for $m$ and the tree-graph geometry, in terms of the communication length $L$. The markers denote: {\protect\tikz \protect\draw[draw=deeppink, thick] (0,0) circle (0.3em) -- (-0.4,0) -- (0.4,0); } optimized rate obtained with our protocol, {\protect\tikz \protect\draw[draw=GCyan, thick] (0,0) circle (0.3em) -- (-0.4,0) -- (0.4,0); } optimized rate obtained with the protocol of Ref.~\cite{HANNES_PRX}, considering symmetric LTCs of depth $3$ and with maximum  branching parameter set to $20$.  
}
\label{fig_plots}
\end{figure*}

\indent The recovery protocol performed at each repeater station has been discussed extensively in previous literature for the case of symmetric tree-graphs~\cite{VARNAVA_LOSS, HANNES_PRX}; there, the success probability $P_\text{rec}$ takes the compact form
\begin{equation}\label{symmetric_Prec}
    P_\text{rec} / (1-\epsilon) = \frac{P_\mathcal{X}}{1-\epsilon} \Big[P_\mathcal{Z}^N - \left(\epsilon P_{\mathcal{Z}_\text{ind}}\right)^N\Big],
\end{equation}
where $\epsilon$ denotes the probability that one single photon is lost in the channel. Therein, $P_\mathcal{Z}$ and $\mathcal{P}_\mathcal{X}$ depend on both $\epsilon$ and the specific choice of LTC; moreover, the relation $P_\mathcal{Z}=1-\epsilon + \epsilon P_{\mathcal{Z}_\text{ind}}$ holds.\\
\indent In Appendix~\ref{appendix success probaility}
we show in detail how the equation above generalizes for generic LTCs, resulting in the following expression for the success probability:
\begin{equation}
	P_\text{rec} / (1-\epsilon) =  \sum_{k=0}^{N-1} \epsilon^{k}P_{\mathcal{X}}(k) \Bigg[\prod_{i=0}^{k-1} P_{\mathcal{Z}_\text{ind}}(i)   \!\prod_{j=k+1}^{N-1} P_{\mathcal{Z}}(j)\Bigg].
	\label{Ps generale}
\end{equation}
We now analyze the parameters herein, leaving the derivation to Appendix~\ref{appendix success probaility}. First, note that we can characterize any generic tree-graph by specifying (i) the number $N$ of branches attached to the root vertex, and (ii) all the tree-sub-graphs identifying each branch, which we label by $k=0,1,\dots,N-1$. A depiction is provided in Fig.~\ref{fig_1}(a). Then, to each of these branches we can associate the probabilities $P_\mathcal{X}(k)$, $P_{\mathcal{Z}_\text{ind}}(k)$ and $P_\mathcal{Z}(k)$, which generalize the terms in~\eqref{symmetric_Prec}. Still, the relation $P_\mathcal{Z}(k)=1-\epsilon+\epsilon P_{\mathcal{Z}_\text{ind}}(k)$ holds, so that we can consider only $P_\mathcal{X}(k)$ and $P_\mathcal{Z}(k)$ to characterize branch $k$. These quantities are calculated starting from the graph structure with the methods in appendix \ref{appendix success probaility}. One can straightforwardly verify that Eqs.~\eqref{Ps generale} and~\eqref{symmetric_Prec} coincide for symmetric tree-graphs, i.e. if the dependences on $k$ are dropped. Thus, any argument in what follows recovers the case of symmetric tree-graphs studied in the previous literature, upon considering the asymmetric case.  \\
\indent We now inspect Eq.~\eqref{Ps generale} to give an intuition as to why asymmetric tree-graphs improve the LTC performance. To this end, we argue as follows. Recall that the goal is to optimize the branches $k=0,1,\dots$ one by one, so to maximize $P_\text{rec}$. From the explicit form in Eq.~\eqref{Ps generale}, we note that the factor $\epsilon^{k}$ generally causes terms with small $k$ to give the most important contributions to the overall summation; this is particularly true for small values of $\epsilon \ll 1$. In view of optimizing the LTC, we therefore heuristically expect that the key requirement for optimal LTCs is that $P_\mathcal{X}(k)$ is high for small $k$, while $P_{\mathcal{Z}s}(k)$ is high for large $k$.\\
\indent Importantly, for a branch $k$, it is \emph{not} possible to optimize the geometry so to maximize \emph{both} $P_\mathcal{X}(k)$ and $P_{\mathcal{Z}}(k)$ simultaneously; more specifically, it turns out that if one wishes $P_\mathcal{X}(k)$ to be high, then $P_{\mathcal{Z}}(k)$ will automatically be penalized, and vice-versa. This is illustrated in Fig.~\ref{fig_plots}(a): there, for generic branch geometries, we display the quantities $1-P_\mathcal{X}$ and $1-P_\mathcal{Z}$. From the resulting plot, it is manifest that indeed they have opposite behavior, in that one is only large when the other is small.\\
\indent An optimization over symmetric tree-graphs, where $P_{\mathcal{X},\mathcal{Z}}(k)=P_{\mathcal{X},\mathcal{Z}}$ are uniform over the branches, thus searches for a compromise that balances this tension; as an example, in the main plot in Fig~\ref{fig_plots}(a) we mark the optimal symmetric tree-graph across the ones considered (i.e., the one that maximizes $P_\text{rec}$) for a specific case-study (see figure caption). In sharp contrast, if we optimize over asymmetric tree-graphs, we can explore configurations where e.g. the branches with small $k$ are of the form which maximizes $P_\mathcal{X}$ (top-left points in the plot), while for large $k$ we can pick branches from the bottom-right part of the plot, which instead maximise $P_\mathcal{Z}$. This suggests that the optimal solution is \emph{not} a symmetric one, where all branches $k$ are equal, but rather one that combines different types of branches, which are judiciously placed for increasing $k$.\\
\indent For these reasons, here we consider generally asymmetric LTCs. We note, however, that optimizing over \emph{all} possible tree-graphs is exponentially more costly than only optimizing over symmetric tree-graphs (specifically, the number of graph-parameters is exponentially larger). Our optimization is thus also based on heuristic methods, where we employ \emph{ans\"atze} for the geometry of each branch based on its position within the tree-graph. More precisely, we first search for the branches which display unbalanced performance towards $P_\mathcal{X}$ or $P_\mathcal{Z}$, and then we feed them as starting ans\"atze to the optimization of the entire LTC. While this approach does not guarantee global optimality, still it is sufficient to  drastically outperform previous approaches based on (globally) optimized asymmetric tree graphs.  \\
\indent As a final note, we remark that, as shown in the inset of Fig.~\ref{fig_plots}(a), the effect of these asymmetries is progressively reduced when one considers larger tree-graphs (e.g., with deeper branches or larger branching parameters). For this reason, previous work focused on relatively large-sized graphs~\cite{VARNAVA_LOSS, HANNES_PRX, ECONOMOU_PERFORMANCE}. However, the graph size impacts dramatically the communication rate, as it increases considerably both the total number of photons required and the complexity involved in generating the LTCs themselves. Instead, our proposed optimization does not require to increase the number of photons; in fact, we actually find that our method generally reduces systematically the number of photons composing the LTCs, and consequently many other important parameters, such as the generation rate.

\begin{figure*}[t!]
    \begin{minipage}{1.0\textwidth}
        \includegraphics[width=\textwidth]{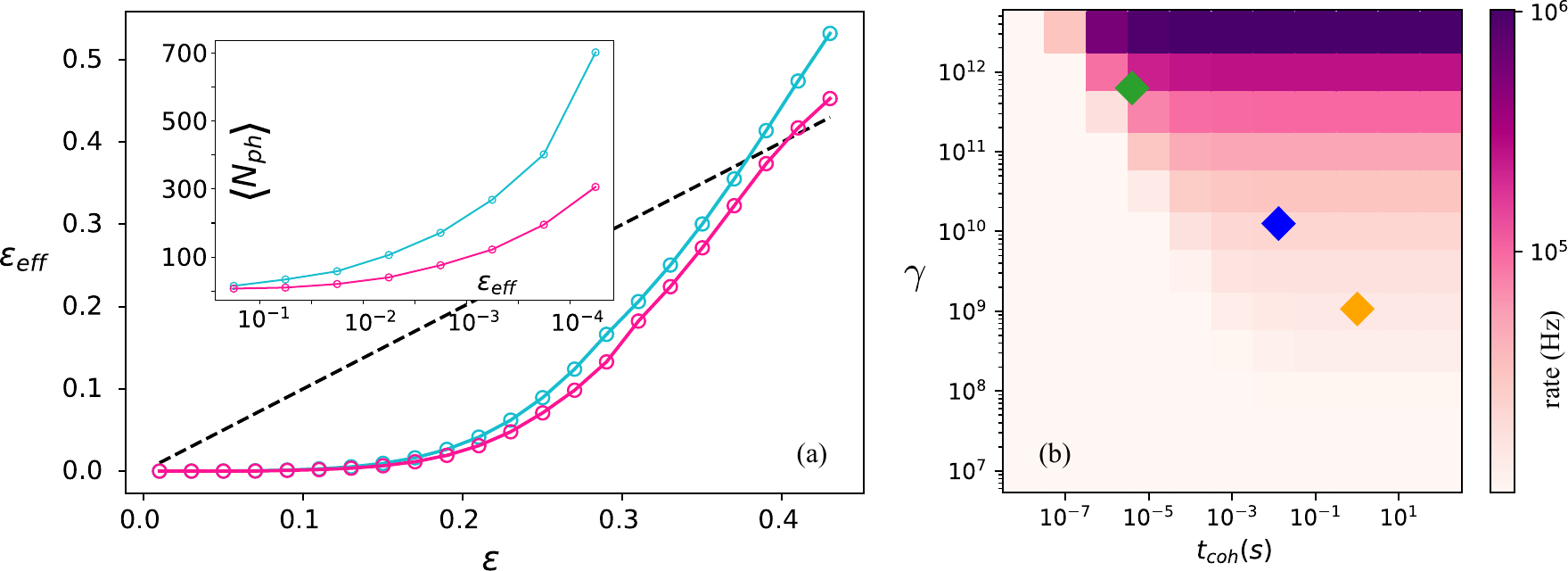}
    \end{minipage}%
\caption{In Fig.(a) every point of the main plot is obtained minimizing $\epsilon_\text{eff}$ as a function the tree-graph geometry. With {\protect\tikz \protect\draw[draw=deeppink, thick] (0,0) circle (0.3em) -- (-0.4,0) -- (0.4,0); } we denote the results for the asymmetric tree-graph case, while  with {\protect\tikz \protect\draw[draw=GCyan, thick] (0,0) circle (0.3em) -- (-0.4,0) -- (0.4,0); } the symmetric one. In both scenarios we take into account LTCs with a maximum number of photons $N_\text{ph} = 100$. In the inset, we fix $\epsilon=0.1$ and plot the number of photons which is needed for achieving a desired effective loss-rate $\epsilon_\text{eff}$; our results highlight how asymmetric LTCs allow significantly lower photon numbers at comparable loss-correction regimes. In Fig.(b) the rate $\mathcal{R}$ as function of the optical line width $\gamma$ and the coherence time of the quantum emitter $t_\text{coh}$, for $L = 100 $ km is reported. The markers correspond to different technological implementations for the emitters. \textcolor{tabgreen}{\ding{117}} a single semiconductor quantum dot coupled with a nano-cavity, \textcolor{blue}{\ding{117}} a single silicon-vacancy (SiV) color center coupled with a photonic crystal cavity and \textcolor{orange}{\ding{117}} a single neutral atom coupled
with a fiber Fabry-Perot cavity. }
\label{fig_6_7}
\end{figure*}

\section{Results: quantum repeater performance}\label{section6}

We now evaluate the employment of our generation protocol, as well as our code optimizations, for quantum repeater schemes. For this, we consider the one-way quantum repeater scheme of Ref.~\cite{HANNES_PRX}, which builds upon the generation methods first introduced in Ref.~\cite{economou_prx} and is entirely based on tree-graph LTCs. 

\subsection{Figures of merit and parameters}

We consider a transmission line of length $L$, which is divided in $m$ smaller parts of length $L/m$ by repeater stations; at each station, the signal is purified from losses (decoding-encoding). Similarly e.g. to Refs.~\cite{HANNES_PRX, ECONOMOU_PERFORMANCE}, we specifically consider the well-known six-state variant of the BB84 protocol for QKD~\cite{scarani2009security}. In the asymptotic limit of infinitely long keys, the secret-bit rate $f$ appearing in Eq.\eqref{rate_formal} takes the form 
\begin{equation}
        f= 1-h(Q) -Q- (1-Q)h \left(\frac{1-3Q/2}{1-Q} \right) ,
    \label{secret key fraction}    
    \end{equation}
where $h(x)=-x\log(x) - (1-x)\log(1-x)$ is the binary Shannon entropy. The parameter $Q$ quantifies the potential operational errors in the encoding-decoding at the repeater stations, as well as decoherence and depolarization of the photonic qubit during the transmission. It is evaluated as $Q=2\epsilon_\text{trans}/3$, where $\epsilon_\text{trans}$ is the probability of an error occurring at any of the $m$ stations; this is calculated as $\epsilon_\text{trans}=1-(1-\epsilon_r)^{m+1}$, where $\epsilon_r$ is the local operational error. Finally, the probability $P_\text{succ}$ in Eq.~\eqref{rate_formal} is calculated as the overall probability that the transmission of a single bit of information succeeds through the communication line; thus, it is given by $P_\text{succ} = P_\text{rec}^{m+1}$, where $P_\text{rec}$ is the loss-corrected recovery probability given by Eq.~\eqref{Ps generale}. The loss probability $\epsilon$ appearing in the expression~\eqref{Ps generale} for $P_\text{rec}$ is defined as $\epsilon = 1-\eta_\text{d} \exp\left(-L_\text{eff}/L_\text{att}\right)$. Herein, $\eta_\text{d}$ is the overall detection efficiency at the repeater stations, $L_\text{att}$ is the attenuation length of the optical fiber, and  $L_\text{eff}$ the physical distance effectively traveled by each photon. The latter quantity is calculated by also considering any delay line needed for the generation of the LTC; thus, it reads $L_\text{eff} = \tau_\text{del}c + L/m $, where $c$ denotes the speed of photons in the waveguide and $\tau_\text{del}$ is the total delay time.

\subsection{Improved quantum repeater rates}
We now elaborate on our numerical analysis. In Fig.~\ref{fig_1}(c) we fix the transmission distance $L=300\mathrm{km}$, and we plot the achieved rate $\mathcal{R}$ as a function of the number of repeater stations $m$. Specifically, for any $m$ we optimize the LTC so to maximize $\mathcal{R}$. These results are already discussed in section~\ref{section2}. Comparing with previous proposals, this showcases how one main advantage of our protocol is that we can achieve considerably higher rates with much fewer repeater stations. \\
\indent The numerical results in Fig.~\ref{fig_1}(c) are obtained as follows. For each $m$, we consider separately our optimized LTCs generated with the hierarchical protocol outlined in Section~\ref{section4}, and the original codes and generation protocols of Refs.~\cite{HANNES_PRX, economou_prx}. For the latter case, the optimization is straightforward and exact due to the limited parameters involved: we search through all possible symmetric tree-graphs of depth up to $K=3$ and with potentially very large branching parameters, limited by a cutoff that we set at $b_k\leq 20$. This allows us to optimize over symmetric LTCs containing up to approximately $10^3$ photons. Instead, for asymmetric LTCs we follow a non exact approach, where we restrict to heuristically convenient geometries as discussed in Section~\ref{section5}; note that this does not necessarily guarantee to find the optimal solution. Moreover, the proposed top-to-bottom generation scheme imposes several restrictions on the achievable tree-graphs. For the optimization loop, among the allowed geometries, we only consider asymmetric tree-graphs with up to $75$ photons, thus orders of magnitude smaller than the accepted symmetric ones. Specifically we take into account asymmetric tree-graphs with a maximum of four branches attached to the root vertex; moreover, a series of requirements arising from the generation protocol further restrict the achievable LTCs. We characterize more in detail the restricted family of asymmetric tree-graphs our optimization runs through in Appendix~\ref{modification to the pilot}.\\
\indent Remarkably, even though the considered asymmetric tree-graphs have a significantly more restricted number of total photons with respect to the considered symmetric ones, our results show that still much higher communication rates are possible. This is in line with expectations on hierarchically generated tree-graphs, which were indeed conjectured to allow better correction rates~\cite{ECONOMOU_PERFORMANCE}. In addition, these results also strongly benefit of the asymmetries introduced in the graphs: indeed, we note that in many cases asymmetries can be understood as simply `removing' vertices from a symmetric graph, thus automatically lowering the photon count. This highlights that previous works based on symmetric LTCs were to all effects penalized by the presence of many photons in the graph, which actually contributed negatively to the loss-correction performance, and whose employment was unavoidable just because of the imposed symmetry. \\
\indent 
Next, in Fig.~\ref{fig_plots}(b) we report the achievable repeater rates $\mathcal{R}$ for varying distances $L$ across transmission lines extending up to $600\: \mathrm{km}$. Again, we compare our results with state-of-the-art methods. For each value of $L$, we now optimize the rate as a function of the number of repeater stations $m$ and the tree-graph geometry. The geometric constraints on the graph are the same as discussed above for Fig.~\ref{fig_1}(c). Interestingly, we note that in the regime considered and under the constraints imposed, the optimizer for asymmetric tree-graphs always chooses the same LTC, which is depicted in the inset and has the form $\bm{B}=\left((4,3), (4,2),(3,1)\right)$; differently, the optimal tree-graph for symmetric LTCs varies significantly with $L$. From the plot, we observe remarkably higher repeater rates for our protocols, with improvements up to an order of magnitude at considerable distances.\\
\indent As we also discuss more in detail in the section below, our improved rates are largely facilitated by the fact that typically our LTCs feature significantly less photons than previously considered symmetric tree-graphs. This is highlighted in the inset of Fig.~\ref{fig_plots} (b), where we plot the number of photons contained in the chosen optimal LTCs. Therein, we can appreciate how symmetric LTCs need very large numbers of photons for achieving modest rates, as opposed to the optimized asymmetric LTC, which maintains a good performance without the need of increasing the photon number.

\subsection{Impact of asymmetric encoding}

To evaluate the impact of our asymmetric optimization, we now isolate the sole effect of employing asymmetric tree-graph structures for LTCs. For this, it is useful to consider the most basic task of loss correction. Specifically, we study the paradigmatic situation where each physical qubit (photon) has a probability $\epsilon$ to be lost, and for a given LTC we compute the logical loss-rate of the error-corrected LTC, 
 \begin{equation}
     P_\text{rec} = f_\text{LTC}(\epsilon).
 \end{equation}
This is calculated as in Eq.~\eqref{Ps generale}. Our results are shown in the main plot of Fig.~\ref{fig_6_7} (a), which is obtained as follows. For both, symmetric and asymmetric tree-graphs, we fix a maximal number of photons in the LTC, that is, we set $N_\text{ph}\leq 100$. For each $\epsilon$ we optimize the tree-graph under the constraint that it never exceeds $N_\text{ph}$ vertices, and we identify the one with the highest $P_\text{rec}$. Note that in the plot we display $\epsilon_\text{eff} = 1-P_\text{rec}$. We thus certify that asymmetric LTCs achieve lower or comparable effective losses, and a higher break-even point than symmetric ones. But most importantly, we find that, typically, to achieve a specific $\epsilon_\text{eff}$, using asymmetric tree-graphs significantly fewer photons per LTC are needed, as shown in the inset plot. In the latter, we now fix a raw loss rate $\epsilon = 0.1$, and we plot the needed number of photons for achieving a desired effective rate $\epsilon_\text{eff}$. In appendix \ref{numerical methods} we explain how this is evaluated numerically. Despite the optimization not covering the entire configuration space of asymmetric tree-graphs, our results show that they always achieve comparable correction with a significant reduction in the total photon number. For instance, for $\epsilon_\text{eff}\leq 10^{-4}$ the asymmetric tree-graphs require less than half the number of photons, compared to the symmetric ones.\\
\indent To summarize, the benefit of using asymmetric LTCs becomes mostly marked when applied to quantum repeater protocols, as they offer two key advantages in improving the communication rate in Eq.\eqref{rate_formal}. First, the reduced photon number directly decreases the LTC generation time, $T_\text{rep}$. Second, as demonstrated in previous plots, they enhance the recovery probability while minimizing resource usage.

\subsection{Impact of hierarchical generation: neutral atom implementation}
\begin{figure}
    \centering
    \includegraphics[width=1\linewidth]{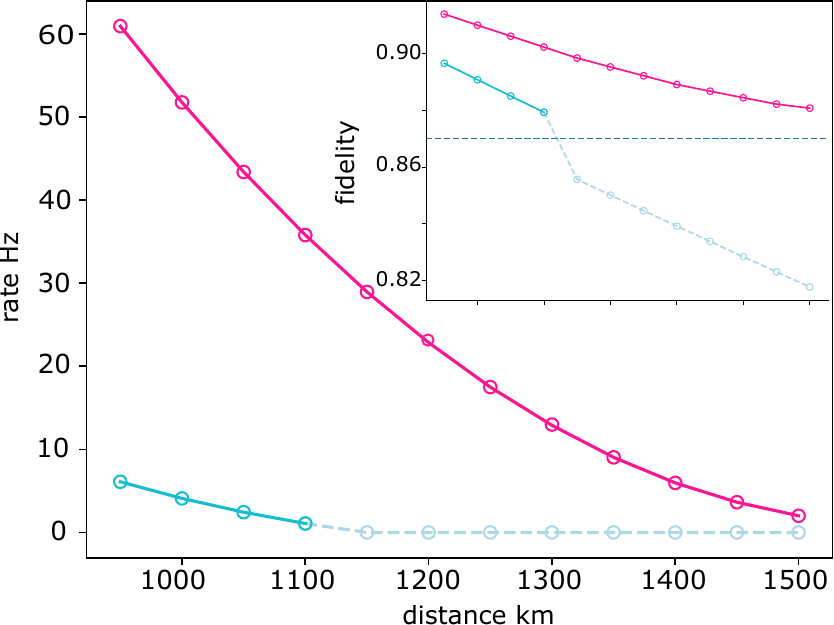}
    \caption{In the main plot we show the optimized rate of Eq.~\eqref{rate_formal}, over all the possible tree geometries generable with our single emitter top-to-bottom protocol, and over the number of repeating station $m$, for different distances. With {\protect\tikz \protect\draw[draw=deeppink, thick] (0,0) circle (0.3em) -- (-0.4,0) -- (0.4,0); } we denote the results for top-to-bottom generation, while  with {\protect\tikz \protect\draw[draw=GCyan, thick] (0,0) circle (0.3em) -- (-0.4,0) -- (0.4,0); } the results for bottom-to-top generation. The points shown with dashed lines and a lighter color correspond to cases where it is not possible to extract any secret key, due to the fidelity being below threshold. In the inset we show the fidelity relative to the optimal configuration that maximizes the key rate in both generation schemes. We also indicate the threshold below which secret key extraction is no longer possible. Here we consider explicitly neutral atom hardware parameters.} 
    \label{tbvsbt}
\end{figure}
We now discuss the advantages of hierarchical generation. The key concept is as in Eq.~\eqref{effective_p_loss}. In our protocols, not reordering the photons after the generation avoids an enhancement of the loss probability (exponential in system size), that would otherwise be enforced~\cite{ECONOMOU_PERFORMANCE}. In practice, this is mostly significant in systems where the emission rate $\gamma$ is low: if subsequent photons are separated in time by a scale $1/\gamma$, very long delay lines (of length $\sim c/\gamma$) must be employed for the recording, resulting in very high losses. A clear example of this are neutral atom systems: here, the long coherence time ($t_\text{coh}\simeq 1 $s, much larger than $1/\gamma\simeq 10^{-9} $s) would in principle allow high-fidelity operations, with the generation of very large LTCs before any error occurring with non negligible probability; but the long emission time, combined with the reordering, results in most of the photons being lost at the repeater station, rather than in the main channel. Indeed, previous work analyzing bottom-to-top schemes has found that neutral atoms, for long distances ($L \gtrsim 1000$km), could not surpass the fundamental secret-bit rate threshold of $f\gtrsim 0.89$, under which no secure communication is possible. In Fig. \ref{tbvsbt} we compare the rate for top-bottom and bottom-top generation with neutral atom parameters. For top-bottom generation we employ our protocol using the methods of Section~\ref{section4}, while for bottom-top generation we consider the methods of Refs.~\cite{HANNES_PRX, economou_prx}; we optimize over the number of stations $m$ and all possible tree-graph LTC geometries generable with our single emitter protocol. In the figure we show that bottom-top generation soon descends below the fidelity threshold, as was also found in Ref.~\cite{ECONOMOU_PERFORMANCE}; below this threshold, it is no longer possible to distill a secret key, resulting in a communication rate $\mathcal{R}=0$. In contrast, our top-to-bottom generation avoids the exponential fidelity suppression of Eq.~\eqref{effective_p_loss}, resulting in a clear fidelity improvement, which directly translates to significantly higher rates and a considerable shift of the threshold point.

\section{Hardware considerations}
\label{hardware considerations}
Several physical systems are well suited for implementing our generation scheme, including quantum dots~\cite{senellart2017high,Uppu_2021}, silicon-vacancy color centers~\cite{Evans_2018} and single atoms in cavities~\cite{RevModPhys.87.1379}. Specific designs are envisioned in Appendix~\ref{specific_implementations}. In Fig.~\ref{fig_6_7}(b) we evaluate the achievable repeater rates in realistic scenarios, considering the dependence on the decay rate $\gamma$, and also taking into account the limited coherence time $t_\text{coh}$ of the emitter. Our calculations span across several regimes to cover this wide range of physical implementations. In the figure, the rate is optimized over a fixed distance of $L=100\mathrm{km}$, and the maximum photon number in a code is constrained to relatively small sizes, $N_\text{ph}  \lesssim 100$~\footnote{To be precise, in the optimization loop we technically consider tree graphs that can be generated with our hierarchical protocol, restricting to four layers. Moreover, we limit the last layer to at most $20$ photons per branch. Together with the natural constraints on the LTCs which can be generated with our protocol, this results in a theoretical maximum photon number of $N_\text{ph}\leq 298$ for the graphs considered in our loop. However, due to the strong penalty given by $N_\text{ph}$ in Eq.~\eqref{rate_formal}, it turns out that our optmization program never chooses graphs with more than $4$ vertices per branch in the last layer. This results in the fact that the relevant search is actually on much smaller graphs, with $N_\text{ph}\leq 74$}. \\
\indent Current technologies are theoretically capable of achieving significant rates with remarkably small LTCs. For instance, by employing as emitter a semiconductor quantum dot coupled to a microcavity, with emission rates in the range $\gamma \sim 2\pi\times 100\mathrm{GHz}$\cite{gamma_qdot} (and coherence times $t_\text{coh}\sim 4\mu\mathrm{s}$\cite{tcoh_qdot}), one can achieve $\mathcal{R}\sim 117.3\mathrm{kHz}$ with a hierarchically emitted asymmetric LTC containing only $N_\text{ph}=35$ photons. In this construction, the number of needed repeater stations is $m=44$, resulting in one every $\sim 2\mathrm{km}$. To fairly compare with other studies~\cite{HANNES_PRX, ECONOMOU_PERFORMANCE}, it is also useful to multiply our cost function by $N_\text{ph}$, so to retrieve the same figure of merit $\mathcal{R}'\equiv N_\text{ph}\mathcal{R}$; in this case, we get $\mathcal{R}'\sim 4.1\mathrm{MHz}$. We also find favorable numbers when the emitter is compatible e.g. with a silicon-vacancy color center coupled to photonic crystal cavities, where smaller emission rates $\gamma\sim 2\pi\times 2\mathrm{GHz}$\cite{bhaskar2020experimental} are well compensated by much longer coherence times in the range $t_\text{coh}\sim 13 \mathrm{ms}$\cite{Silicon_vacancy_coherece}; here, we find an achievable rate in the range of $\mathcal{R}' \sim136.5\mathrm{kHz}$ with repeaters positioned every $\sim1.75\mathrm{km}$, and by employing LTC with only $N_\text{ph}=32$ photons.\\
\indent We finally highlight that neutral atoms in Fabri-Perot cavities, which feature remarkably high coherence times in the range $t_\text{coh}\sim 1\mathrm{s}$\cite{tcoh_atomo}, can also achieve valuable rates $\mathcal{R}'\sim 9.59\mathrm{kHz}$ with a similar repeater spacing. We remark that the latter fact is in contrast e.g. with the results in Ref.~\cite{ECONOMOU_PERFORMANCE}, where standard bottom-to-top generation methods were employed; therein, this same setup was found to be impracticable for feedback-assisted one-way protocols - even if the tree-graphs considered were much larger. Indeed, there the main limitation was the slow emission of neutral-atom systems, with rates in the range of $\gamma \sim 2\pi\times 170\:\mathrm{MHz}$ \cite{gamma_atomo}; due to the consequently large time-spacing between sequentially emitted photons, this resulted in the need for extremely long delay lines for implementing both the feedback loop and the photon reordering, eventually introducing more losses than the LTC could correct. In contrast, here our methods allow to both avoid the reordering and to drastically reduce the number of fed-back photons, resulting in considerably reduced loss during the decoding-encoding procedure, and thereby allowing for efficient error-correction.

\section{Conclusion}\label{conclusion}
In summary, we have proposed and analyzed a novel class of protocols for efficiently preparing LTCs, focusing on their integration in quantum repeater schemes. As opposed to current proposals~\cite{economou_prx, HANNES_PRX, ZHAN_PRL,ECONOMOU_PERFORMANCE}, with our method the generation proceeds hierarchically, i.e. the error-correcting code (the tree-graph state) is prepared from top to bottom; this allows substantial savings at the repeater stations, e.g. by avoiding to reorder the photons during the recovery. Importantly, we showed that this allows for much shorter delay lines for implementing the feedback loop. Together, these advances significantly reduce the photon loss during the decoding-encoding procedure at the repeater stations. In addition, we have also shown that, contrary to the main belief, loss correction is enhanced by \emph{asymmetric} encodings, specially at low code sizes - a direct effect of the intrinsic asymmetry in the `counterfactual' recovery protocol~\cite{VARNAVA_LOSS, HANNES_PRX}. Our setup employs a static hardware layout, where the entire dynamics is regulated by just controlling the laser driving on a single emitter, which is subjected to a simple time-delayed feedback mechanism; this can be realized e.g. by employing a mirror or chiral couplings between the emitter and the waveguide. \\
\indent In agreement with the main intuition, numerical analyses show that our method is capable of competitive loss correction and high-rate communication with significantly lower photon numbers per error-correcting code; this suggests that near-term implementations might especially benefit from our results. Emitters such as quantum dots~\cite{Uppu_2021} and silicon-vacancy~\cite{Evans_2018} centers are suitable candidates for implementing our scheme, allowing for high secret bit rates across hundreds of kilometers and with minimal resources, by employing loss-tolerant quantum codes featuring few tens of photons and repeater stations at typical distances of $\sim2\mathrm{km}$. We have also discussed that, in contrast with similar proposals~\cite{ECONOMOU_PERFORMANCE}, our methods are also well suited for neutral atoms in cavities~\cite{RevModPhys.87.1379}, as (because of the relatively low emission rate in the $\mathrm{MHz}$ regime) they especially benefit from the loss-saving in our hierarchical protocols. \\
\indent We remark that state of the art platforms are now closing the gap towards the hardware requirements necessary for implementing our scheme, with improved emitter-waveguide cooperativities~\cite{Najer_2019, Uppu_2020} and important advances in emitter-photon and (emitter-mediated) photon-photon gate engineering~\cite{Liu_2024, staunstrup2024direct}. Building on the simplified setup considered here, together with the much lower photon numbers required, our work potentially opens the door for near-term demonstrations of scalable quantum repeater schemes, and paves the way for future loss-tolerant quantum communications.

\subsection*{Data availability}
\indent Codes and data used for this work are available upon reasonable request via the Zenodo record~\cite{Feri_Zenodo_17226022_2025}.

\section*{Acknowledgments} 
We are grateful for valuable insights from Matteo Marinelli, Hannes Pichler, Francesco Scazza and Alessandro Zavatta. A.B. acknowledges financial support from the PNRR PE National Quantum Science and Technology Institute (GA no. PE0000023), the University of Trieste and INFN. F.C. also acknowledges the financial support
of PON Ricerca e Innovazione 2014-2020 (D.M. 1061,
10.08.21) and QTI (Quantum Telecommunications Italy).

\newpage

\appendix

\section{Physical implementations}\label{specific_implementations}

Here, we elaborate on concrete physical implementations of our proposed generation protocols. 

\subsection{Neutral atom in a Fabri-Perot cavity}

For neutral atoms in cavities, as demonstrated in experiments~\cite{Reiserer_2014, Tiecke_2014, tcoh_atomo}, we envision using the standard hyperfine qubits to realize the states $\ket{g_0}$ and $\ket{g_1}$, while a third ground-manifold (hyperfine/Zeeman) sublevel can be used for $\ket{g_2}$ to enable slow, narrowband Raman emission when required. For the excited state $\ket{e}$, one can use the $D-$line. For concreteness, considering the standard example of $^{87}\mathrm{Rb}$, a natural choice is to encode the ground-state qubit through the clock pair $\ket{g_0}\equiv\ket{5S_{1/2},F=1, m_F=0}$ and $\ket{g_1}\equiv \ket{5S_{1/2},F=2, m_F=0}$, while for the excited state we can use the $D_2$ line via $\ket{e}\equiv \ket{5P_{3/2}, F'=3, m'_F=1}$. A sound choice for the intermediate state can be $\ket{g_2}\equiv \ket{5S_{1/2},F=1, m_F = +1}$. \\
\indent We note that the choice of the $m_F=0$ clock states for the ground-state qubit are strongly motivated by first-order Zeeman insensitivity, which makes them robust e.g. against magnetic field noise. Crucially, the $\simeq 6.8\mathrm{GHz}$ hyperfine gap separates the two levels enough, such that scattering photons can be considered off-resonant with respect to $\ket{g_0}$, and the scattering phase necessary for our entangling gates is only acquired by $\ket{g_1}$, as desired. Finally, choosing a different manifold $F$ for $\ket{g_2}$ allows a well-resolved $\Lambda$ system for the slow emission protocols.  

\subsection{Semiconductor quantum dots}

We encode the matter qubit in the ground electron-spin Zeeman states and use a spin-selective trion transition for the conditional scattering; a shelved (dressed) ground state provides the auxiliary level for Raman emission. Specifically, we set $\ket{g_0} \equiv \ket{\uparrow}$, $\ket{g_1} \equiv \ket{\downarrow}$. For the excited state, we can encode it as $\ket{e} \equiv \ket{T_\downarrow}$, choosing a spin-selective trion addressed from $\ket{\downarrow}$. For $\ket{g_2}$, we can use an auxiliary long-lived dressed/sheld=ved ground-manifold state, forming a $\Lambda$ system with $\ket{e}$. \\
\indent Here the feedback photon is resonant with the cavity-enhanced $\ket{g_1}\!\leftrightarrow\!\ket{e}$ transition, yielding the conditional phase on reflection. When slow, narrowband emission is required, we drive a Raman process on the $\ket{g_2}\!\rightarrow\!\ket{e}$ leg; all entangling scatterings still address $\ket{g_1}\!\leftrightarrow\!\ket{e}$. (Faraday geometry with circular polarization gives the usual selection rules.)

\subsection{Silicon-vacancy centers in diamond (SiV$^-$ in a nanocavity/waveguide).}
We use two spin sublevels in the lower orbital branch of the ground ${}^{2}E_g$ manifold, such that the ground qubit is encoded through $m_s = \pm 1/2$. For $\ket{e}$, we address an optical line into ${}^{2}E_u$, providing a spin-/orbit-selective line from $\ket{g_1}$. Finally, an upper-branch ground sublevel acts as the shelving state, thereby encoding $\ket{g_2}$.  \\
\indent To implement our protocol, the feedback photon is tuned to the $\ket{g_1}\!\leftrightarrow\!\ket{e}$ transition, realizing the state-dependent phase, while the $\Lambda$ system $\ket{g_2}\!\leftrightarrow\!\ket{e}\!\leftrightarrow\!\ket{g_1}$ enables slow, narrowband Raman emission when needed.

\section{Modifications to the pilot protocol}
\label{modification to the pilot}
\begin{figure}
        \centering
        \includegraphics[width=1\linewidth]{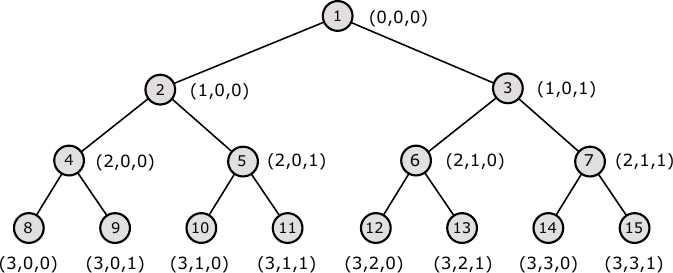}
        \caption{Tree graph state generated using the non-optimized version of the pilot protocol, whose emission scheme is illustrated in Fig.~\ref{fig_pilot}. Here, the numbers inside the circled vertices denote the photon emission order, and the labels on the side indicate the indices $(k,p,a)$ specify the qubit position within the tree, as explained in the main text.}
        \label{tree_scheme}
    \end{figure}

\indent Fig.~\ref{tree_scheme} shows the exact emission ordering of photons in our pilot protocol. As we anticipated, there are several modifications that we can make to the pilot protocol. These have essentially two purposes: First, to generate a wider class of LTSs, i.e. to go beyond the simplest case of binary tree-graphs; Second, to speed up the generation, i.e., to generate LTSs at a higher rate. The first purpose is motivated by the fact that, in general, non binary trees can achieve higher performances in loss correction. The second is mostly relevant when the LTS is employed for loss-tolerant quantum communications: indeed, in this case an important figure of merit is the \emph{communication rate}, which is clearly affected in first place by the rate of the LTS generation; for instance, the efficiency of quantum repeater schemes strongly depends on the rate of the local operations at the intermediate repeater stations.  
\begin{figure*}[t!]
    \begin{minipage}{1.0\textwidth}
        \includegraphics[width=\textwidth]{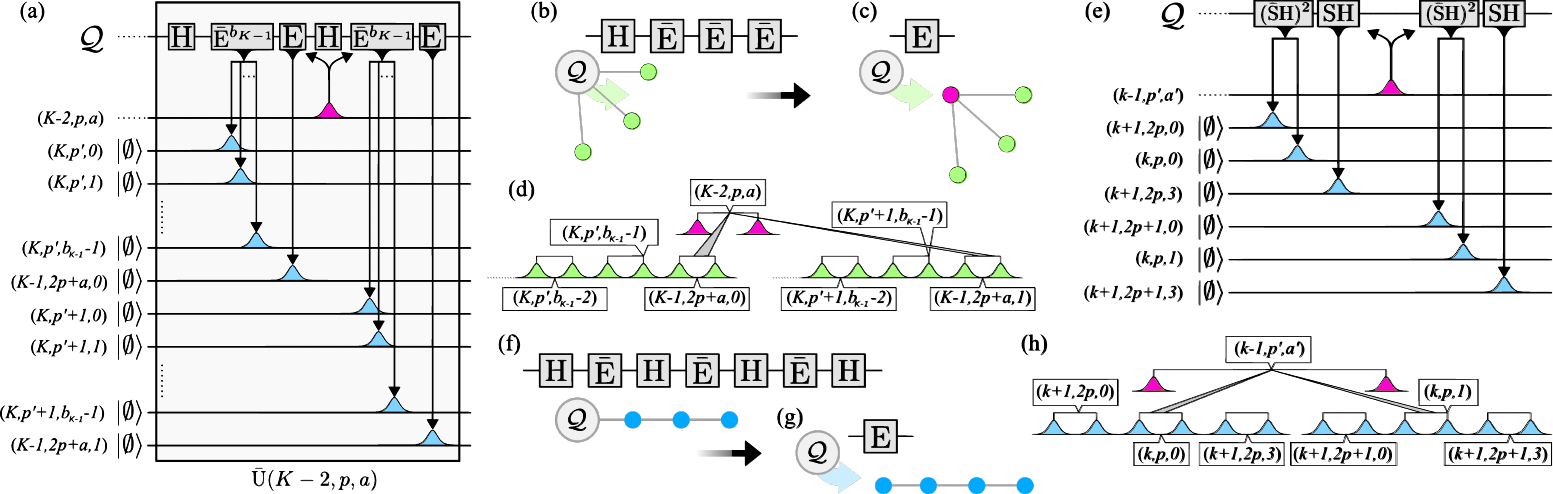}
    \end{minipage}
\caption{(a) Modification 1 to the pilot protocol, which enables to generate LTSs with arbitrary last branching parameter $b_{K-1}$, while also speeding up the generation rate by $\sim 50\%$. (b) The machine-gun driving in Eq.~\eqref{machine_gun_fast_pilot} prepares a GHZ state between $\mathcal{Q}$ and the emitted qubits; (c) a final $\mathrm{E}$ emission disentangles $\mathrm{Q}$ and leaves an all-photonic GHZ state. (d) The timing of the feedback of layer $K-2$ allowing modification 1 to work. (f) The machine-gun driving in Eq.~\eqref{cluster_machine-gun} prepares a linear graph state with $\mathcal{Q}$ at one extreme; (g) again $\mathcal{Q}$ is disentangled. (e) Circuit representation of modification 2; specifically, we show the example of $b_k = 4$. (h) The timing for modification 2.}
\label{fig_modificationa}
\end{figure*}
    \subsubsection{Modification 1 }
    The first modification enables to raise the last branching parameter, $b_{K-1}$, to any arbitrary number, and comes with two main improvements to the pilot protocol: on the one hand, it allows to enhance the loss correction capability by generating a larger class of trees; on the other hand, it drastically lowers the amount of feedback needed (by $\sim 50 \%$), thus enhancing the performance also by reducing the additional losses in the delay line. In addition, we will see that also the LTS generation rate is strongly enhanced, leading to more favorable repeater rates when the protocol is employed for quantum repeater schemes.\\
    \indent Let $b_{K-1}$ be the desired last branching parameter. Then, after the emission of level $K-2$ has been completed, we proceed as shown in the circuit in Fig.~\ref{fig_modificationa}(a). Instead of continuing with the pilot protocol, now we switch from a machine-gun emission like in Eq.~\eqref{pilot_machine_gun} to the following:
    \begin{equation}\label{machine_gun_fast_pilot}
        ...\mathrm{H}\circ \bar{\mathrm{E}}^{b_{K-1}} \circ \mathrm{E} \circ \mathrm{H}\circ \bar{\mathrm{E}}^{b_{K-1}} \circ \mathrm{E} \circ ...,
    \end{equation}
    where essentially we employ several fast emissions like $\mathrm{E}$ and $\bar{\mathrm{E}}$ in place of one slow emission $\mathrm{S}$. This can be understood as follows. As usual, the Hadamard $\mathrm{H}$ prepares the emitter in the superposition $\ket{+}_\mathcal{Q}$; this can be regarded as the preparation $\mathcal{P}$ for emitting a qubit of layer $K-1$ [say, $(K-1,2p+a,a')$, such that $(K-2,p,a)$ is its parent]. However, before emitting $(K-1,2p+a,a')$, we quickly apply $\bar{\mathrm{E}}$ for $b_{K-1}$ times, in order to actually emit its \emph{sons}. Note that, since we are employing $\Bar{\mathrm{E}}$, during this operation $\mathcal{Q}$ remains entangled with all these qubits, never being reset; this is shown in Fig.~\ref{fig_modificationa}(b). Finally, we employ $\mathrm{E}$ for emitting $(K-1,2p+a,a')$, also resetting $\mathcal{Q}$ [see Fig.~\ref{fig_modificationa}(c)]. \\
    \indent During the sequence~\eqref{machine_gun_fast_pilot}, layer $K-2$ is fed-back as in the pilot protocol; the timing is set in such a way that each fed-back qubit gets indeed entangled with its children. This is achieved by the timing displayed in Fig.~\ref{fig_modificationa}(d): (i) the feedback of the early time-bin of $(K-2,p,a)$ is nested in between the two bins of $(K-1, 2p+a,0)$, and (ii) the late time-bin of $(K-2,p,a)$ is fed-back to $\mathcal{Q}$ right before the emission of $(K,4p+2a+1,0)$, i.e., the first son of $(K-1, 2p+a,1)$. This timing seems counterintuitive, as it suggests that $(K-2,p,a)$ undergoes a $\mathrm{CZ}$ with $(K,4p+2a+1,0)$; however, it is equivalent to our primary goal of entangling it with $(K-1, 2p+a, 1)$ through a $\mathrm{CZ}$. \\  
    \indent Formally, we can describe this modification by introducing the unitary
    \begin{align}\label{U_bar}
         \bar{\rr{U}}(k,p,a) = & \; \rr{E}^\star_{(k, p, a), (k+1,2p+a,1)} \bar{\rr{E}}^{b_{K-1}} \rr{H}_\cl{Q} \times \nonumber \\
         & \times \rr{E}^*_{(k, p, a), (k+1,2p+a,0)} \bar{\rr{E}}^{b_{K-1}} \rr{H}_\cl{Q}, 
    \end{align}
    which describes the feedback of $(k,p,a)$ while the operation detailed above is executed [Fig~\ref{fig_modificationa}(a)]; we are specifically interested in $\bar{\rr{U}}(K-2,p,a)$. The modified protocol then reads
    \begin{equation}
        \rr{U}_\text{fast}\!=\! \left[\!\prod_{p=0}^{2^{K\!-\!2}\!-\!1}\!\prod_{a\!=\!0}^1\bar{\rr{U}}(K\!-\!2,p,a)\right]\prod_{k=0}^{K-3} \prod_{p=0}^{2^{k-1}-1} \prod_{a=0}^{1} \rr{U}(k,p,a)
    \end{equation}
    in place of Eq.~\eqref{pilot_unitary}.
    Applying this unitary generates a state which is equivalent to $\ket{\Phi_G}$ up to a local change of basis in the last layer qubits:
    \begin{equation}
        \rr{U}_\text{fast} \ket{g_1}\otimes\ket{\emptyset}_\text{tree} = \ket{g_1}\otimes\left[ \bigotimes_{p=0}^{2^{K-1}}\bigotimes_{a=0}^1 \rr{H}_{(K,p,a)} \right] \ket{\Phi_G}_\text{tree};
    \end{equation}
    this local change of basis does not affect the loss-correction capabilities, which are ensured by the local equivalence with $\ket{\Phi_G}$. Note that, on top of this local change of basis, the equation above is still intended to hold up to local $\mathrm{Z}$ corrections. \\
    \indent One important improvement coming with this modification is that now also level $K-1$ must not be fed-back to the emitter; this allows to essentially halve the length $L_\text{delay}$ of the delay line with respect to the original pilot protocol, thus drastically reducing the additional losses during the generation. \\
    \indent We finally remark that this modification, while bringing several improvements discussed above, does \emph{not} add any complexity to the protocol: in fact, there is no price to pay for executing this modified version of the pilot protocol in place of the original. Thus, from now on we will always study the pilot protocol in this modified version.

    \subsubsection{Modification 2}
    We are now interested in modifying the pilot protocol in such a way to allow the generation of larger classes of LTSs, specially non binary tree-graphs. It is indeed widely known that, in principle, optimal performances can be achieved by more general trees $\bm{b}$ rather than the binary ones considered so far. For top-to-bottom generation, generic trees are clearly harder to generate; however, here we show that some simple modifications to the pilot protocol enable to generate a wide class of LTS from top-to-bottom, and still \emph{efficiently}. While other modifications might be possible, here we only focus on those which are reasonably not harder than the pilot protocol itself.\\
    \indent The idea here is strictly related to the linear graph-state generation presented in~\cite{lindner2009proposal}. Specifically, consider now a machine-gun protocol of the form
    \begin{equation}\label{cluster_machine-gun}
        ...\mathrm{H}\circ\bar{\mathrm{E}}\circ \mathrm{H}\circ\bar{\mathrm{E}}\circ \mathrm{H}\circ\bar{\mathrm{E}}\circ \mathrm{H} ...,
    \end{equation}
    with the emitter initially prepared in $\ket{g_1}$. Then, the output of this driving is a \emph{linear} graph state, with the emitter $\mathcal{Q}$ at one extreme. More precisely, if we define $G_{\text{lin}(a,b,...)}$ to be the linear graph with support on the vertices $\left\{a,b,...\right\}$ in this order, one has 
    \begin{equation}
        \rr{H}_\cl{Q} \big(\Bar{\rr{E}}\rr{H}_\cl{Q}\big)^N \ket{g_1}_\cl{Q}\bigotimes_{k=1}^N \ket{\emptyset}_{q_k} = \ket{\Phi}_{G_{\text{lin}}(\cl{Q}, \bm{q})},
    \end{equation}
    with $\bm{q}=\left\{q_1,...,q_N\right\}$, and therefore
    \begin{equation}
        \ket{\Phi}_{G_{\text{lin}}(\cl{Q}, \bm{q})} = \Big[\rr{CZ}_{\cl{Q}, q_N}\prod_{k=1}^{N-1}\rr{CZ}_{k,k+1}\Big]\ket{+}_\cl{Q}\bigotimes_{k=1}^N\ket{+}_{q_k}
    \end{equation}
    according to the definition~\eqref{graph_state_definition}; this is displayed in Fig.~\ref{fig_modificationa}(f). In addition, we can prepare an all-photonic linear graph state by ending the machine-gun above with an $\mathrm{E}$ sequence instead, thus disentangling $\mathcal{Q}$ from the rest; this therefore reads
    \begin{equation}\label{linear_cluster_all_photonic}
        \rr{E}\rr{H}_\cl{Q} \big(\Bar{\rr{E}}\rr{H}_\cl{Q}\big)^{N-1} \ket{g_1}_\cl{Q}\bigotimes_{k=1}^N \ket{\emptyset}_{q_k} = \ket{g_1}_\cl{Q}\otimes\ket{\Phi}_{G_{\text{lin}}(\bm{q})},
    \end{equation}
    such that the linear graph state only has support on the photonic degrees of freedom [see Fig.~\ref{fig_modificationa}(g)]. We now show that, developing on this new building block, we can indeed enhance the branching of the LTS. \\
    \indent To explain the idea, suppose we target $b_{k} = 4$ for some layer $k$. Then, we proceed as follows; also refer to Fig.~\ref{fig_modificationa}(e) for a circuit representation. After the emission of layer $k-1$ has been completed, in place of the driving~\eqref{pilot_machine_gun} for layer $k$, we proceed with 
    \begin{equation}
        ...\circ (\mathrm{S}\mathrm{H})(\Bar{\mathrm{S}}\mathrm{H})^2 \circ (\mathrm{S}\mathrm{H})(\Bar{\mathrm{S}}\mathrm{H})^2 \circ (\mathrm{S}\mathrm{H})(\Bar{\mathrm{S}}\mathrm{H})^2\circ ...,
    \end{equation}
    which is essentially the repetition of the term $\mathrm{S}\mathrm{H}(\bar{\mathrm{S}\mathrm{H}})^2$, i.e. the one in Eq.~\eqref{linear_cluster_all_photonic} with $N=3$. Thus, the output of this term is an all-photonic linear graph state featuring three time-bin qubits. We shall regard these qubits to be $(k+1,2p+a,0)$, $(k,p,a)$ and $(k+1,2p+a,3)$ respectively, in this order; i.e., the central one is the parent, which is part of layer $k$, and the two others are two of its $b_k=4$ children, part of layer $k+1$. One can indeed check that the entanglement connectivity is the desired one. Then, the timing of the feedback of layer $k-1$ is set as displayed in Fig.~\ref{fig_modificationa}(h): this allows to entangle the parent with its children as desired, while not generating unwanted entanglement with the qubits of layer $k+1$. 
    
    \subsubsection{Optimizing the generation}
    \begin{figure}[t!]
        \centering
        \includegraphics[width=1.0\linewidth]{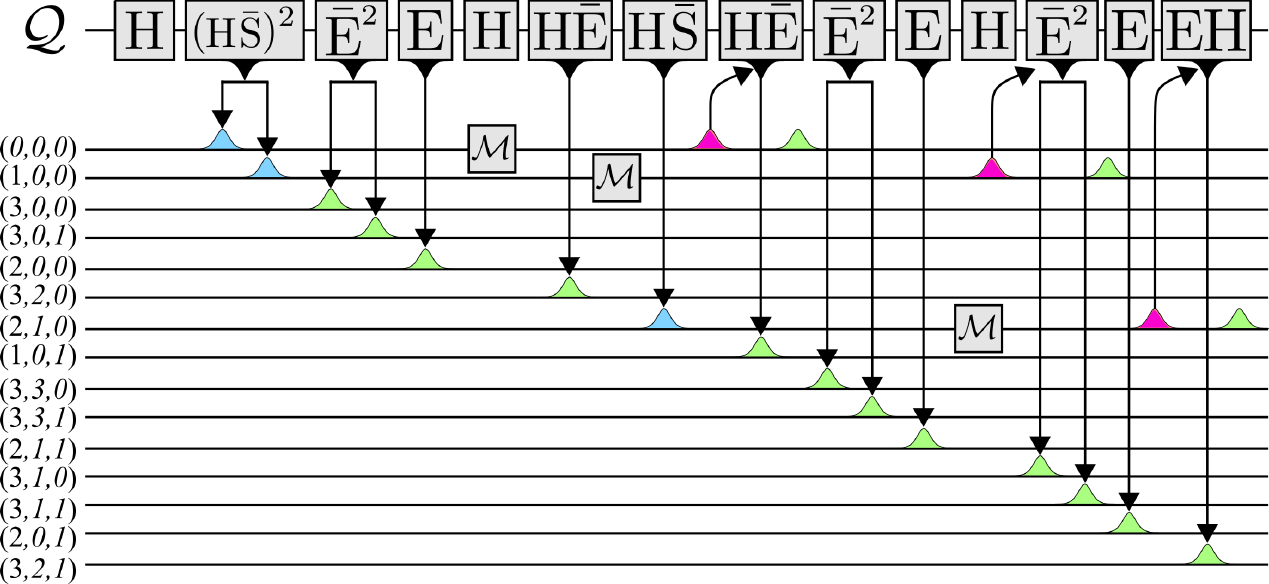}
        \caption{An example of optimized generation of the LTS $\bm{b}=\left\{2,2,2\right\}$.  }
        \label{fig6}
    \end{figure}
    The above ideas can be combined to generate more general graph states and optimize tree-graph generation. This is achieved by judiciously applying the two machine-gun drivings in Eq.~\eqref{machine_gun_fast_pilot} and Eq.~\eqref{cluster_machine-gun}. 
    Here we report the specific driving proposed to generate the trees that we use in the rate optimization. With this generation, we can emit trees with four branches, two of them can be of the type $\bm{b} = \{4,n\}$ and the other two of the type $\bm{b} = \{3,n\}$, where $n$ can be arbitrary.
    \begin{figure}
        \centering
        \includegraphics[width=1\linewidth]{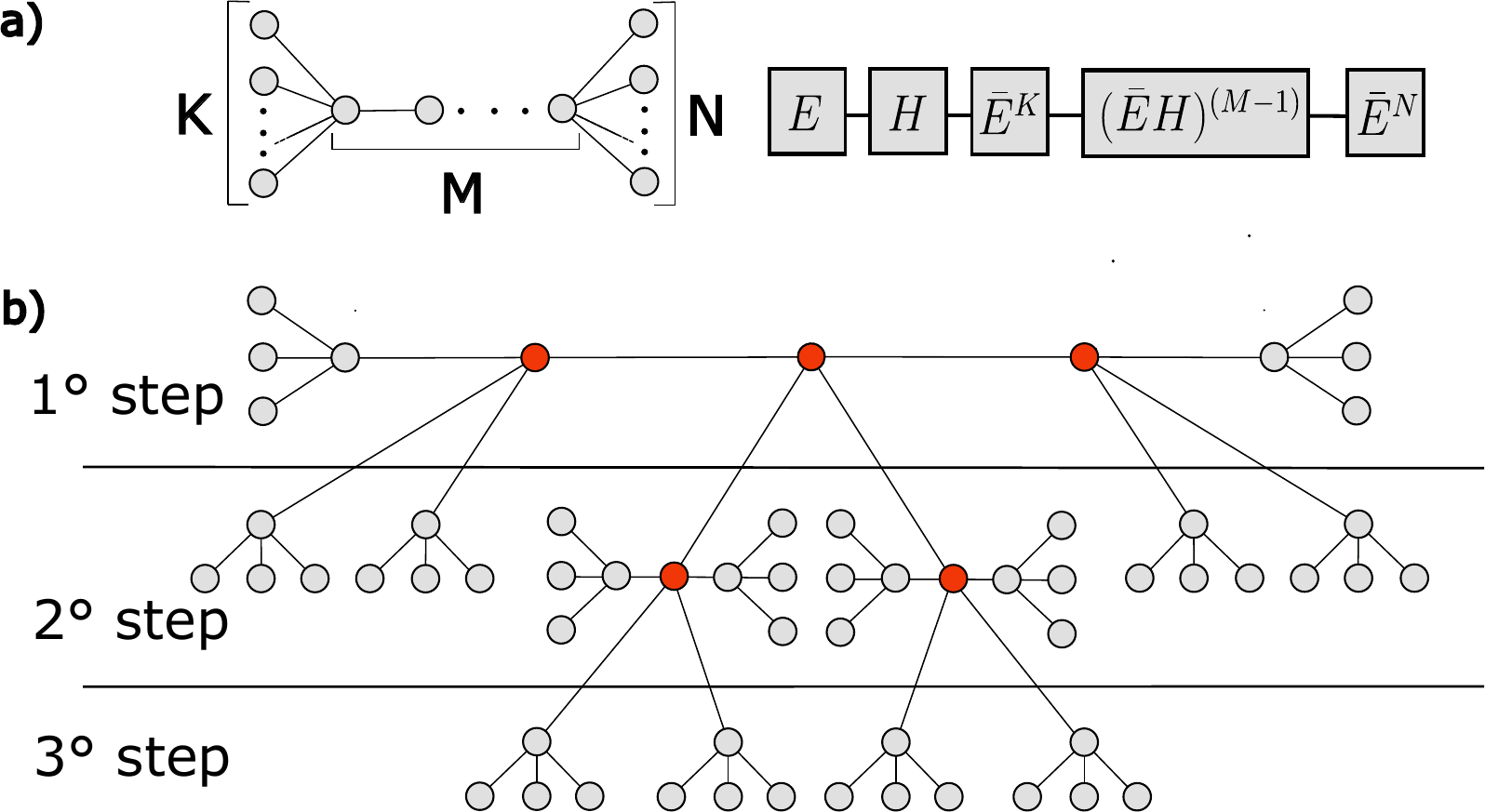}
        \caption{The 3 steps for the top bottom emission of a  general depth 3 tree-graph state, with a single quantum emitter. }
        \label{generation}
    \end{figure}
The fundamental building block of our generation scheme is the graph state, illustrated in Fig. \ref{generation}(a). This state can be generated by following the emission sequence shown on the right side of the figure. A general tree cluster state can then be created by connecting these building blocks using the feedback implemented CZ gates. In Fig. \ref{generation}(b), the three steps needed to emit the cluster state are shown, the photons highlighted in red are the ones that need to be emitted with the slow emission process, in order to ensure the successful interaction between the photon and the emitter during the feedback process.
The length of the delay line must be calculated such that all the photons generated in the longest of the two first steps can fit within it. 
Defining $\tau_n$ the time needed for the emissions in step $n$, the delay time $\tau_{del}$ must be:
\begin{equation}
    \tau_{del} = max(\tau_1, \tau_2).
\end{equation}
The total emission time of the tree-graph $T_{tree}$ is then given by:
\begin{equation}
     T_{tree} = 2\tau_{del} + \tau_2 + \tau_3.
\end{equation}
In conclusion, the described method enables the efficient top-bottom generation of depth-3 tree-graph states, following the branch limitations outlined above, thus optimizing emission times and minimizing the required delay line.

\section{Recovery protocol}
\label{appendix success probaility}
Here we show how Eq. \eqref{symmetric_Prec} for the success probability of the recovery protocol is obtained, and how it generalizes to the case of general asymmetric tree-graphs i.e. Eq. \eqref{Ps generale}.

\subsection{Symmetric case}
Given a symmetric tree-graph, made by $N$ branches that share the same geometry and are described by the branching vector $\bm{b} = (b_1,...,b_n)$, the probability of successfully recovering the information after losses occur, with a certain loss probability $\epsilon$ is given by:
\begin{equation}
    P_{s}= \sum_{k=1}^{N} R_0^{k-1}\epsilon^{k-1} (1-\epsilon)(1-\epsilon+\epsilon R_1)^{b_1} (1-\epsilon+\epsilon R_0)^{N-k },
    \label{p sim}
\end{equation}
where $R_m$ (Eq. \eqref{definizione R_m}) is the probability of performing an indirect $Z$ measurement in the $m$-th level of the tree.
\begin{align}
    & R_m = 1-[1-(1-\epsilon)(1-\epsilon+\epsilon R_{m+2})^{b_{m+1}}]^{b_m}, \nonumber \\
    & R_{m+1}\equiv 0 \qquad b_{m+1} \equiv 0.
    \label{definizione R_m}
\end{align}  
We recall that in order to recover the information the goal is to find a branch where the root qubit is not lost, and then detach this root qubit from all the others, i.e. its children qubits and the root of the other branches, by means of direct or indirect $Z$ measurements. The first tree pieces of Eq. \eqref{p sim}, $R_0^{k-1}\epsilon^{k-1} (1-\epsilon)$, give the probability of losing all the first $k-1$ roots, measure them indirectly along $Z$ (in order to detach them from the rest of the tree) and getting instead the $k$-th root, the fourth piece $(1-\epsilon+\epsilon R_1)^{b_1}$ gives the probability of detaching the root that was not lost from all its $b_1$ children qubits by measuring them along $Z$ directly or indirectly, and the final piece $(1-\epsilon+\epsilon R_0)^{N-k }$ gives the probability of detaching the remaining $N-k$ branches by means of direct or indirect $Z$ measurements of their roots.
In this case where we are dealing with symmetric trees, because all branches share the same geometry, we can evaluate the sum of Eq. \eqref{p sim}:
\begin{equation}
    \label{p sim sommata}
    P_s = [(1-\epsilon+\epsilon R_0)^{N}-(\epsilon R_0)^{N}](1-\epsilon + \epsilon R_1)^{b_1}.
\end{equation}\\
In Eq.\eqref{p sim sommata} only the branching vector element $b_0$ and $b_1$ appear, however, the dependencies from all the other branching vector elements are enclosed in $R_0$ and $R_1$. By defining $P_\mathcal{X} = (1-\epsilon + \epsilon R_1)^{b_1}$, $P_\mathcal{Z} = (1-\epsilon+\epsilon R_0)$ and $P_\mathcal{Z_\text{ind}}= R_0$, we recover the expression present in the main text Eq.\eqref{symmetric_Prec}.\\
Since so far we provided only the definition of $R_m$ in Eq. \eqref{definizione R_m}, here we report how to derive this expression.
$R_m$ as stated, is the probability of measuring a qubit in the $m$-th level along Z indirectly. The requirement for doing this indirect measure is to measure at least one of its children qubits ($m+1$ level) directly along $X$ and all the children qubits of the latter ($m+2$ level) along $Z$ directly or indirectly. Starting from the bottom, the probability of measuring a qubit in $Z$ directly or indirectly in the $m+2$ level is $1-\epsilon+\epsilon R_{m+2}$. Each qubit in the $m+1$ level has $b_{m+1}$ children qubits, and hence the probability of measuring along $Z$ all the qubits of a branch in the $m+2$ level is :
\begin{equation}
      (1-\epsilon+\epsilon R_{m+2})^{b_{m+1}}.
\end{equation}
The probability of not losing a given qubit in the $m+1$ level and measuring all its children qubit in $Z$ is then given by:
\begin{equation}
       (1-\epsilon)(1-\epsilon+\epsilon R_{m+2})^{b_{m+1}}.
\end{equation}
We want the probability that at least one of the $b_m$ qubits of the $m+1$ level is measured in $X$ and all its children qubits are measured in $Z$; to compute this we first calculate the probability that none of the qubits satisfies this requirement that is:
\begin{equation}
    [1-(1-\epsilon)(1-\epsilon +\epsilon R_{m+2})^{b_{m+1}}] ^{b_m},
\end{equation}
and then taking the inverse, we arrive precisely at Eq. \eqref{definizione R_m}.\\

\subsection{Asymmetric case}
We now look at the case of a general asymmetric tree-graph, i.e. a tree-graph where the $N$ branches can have different geometries. The recovery scheme is the same, and the expression for $P_s$ can be extended by adding a index that labels the different branches, becoming:
\begin{align}
    P_s &= \sum_{k=1}^{N}  \prod_{i=1}^{k-1}R_0^{(i)}(1-\epsilon)\epsilon^{k-1}(1-\epsilon+\epsilon R_1^{(k)})^{b_1^{(k)}} \\
    &\prod^{N}_{j=k+1}(1-\epsilon+\epsilon R_0 ^{(j)} ) ,
    \label{Ps genrale 3}
\end{align}
were, $\bm{b^{(k)}}$ is the branching vector describing the $k$-th branch, and consequently $R_m^{(k)}$ is the probability of performing an indirect $Z$ measurement on the $m$-th level of the $k$-th branch. In this case, unlike before, since the branches are different one from each other, we can not further simplify this expression. By introducing the definition for $P_\mathcal{X}$, $P_\mathcal{Z}$ and $P_\mathcal{Z_\text{ind}}$ we recover the expression present in the main text Eq. \eqref{Ps generale}

\section{Numerical methods}
\label{numerical methods}
\subsection{Numerical optimization of asymmetric tree-graphs}
\label{asymmetric generation}
The set of possible asymmetric tree-graphs, made by combining $N$ symmetric branches, with maximum branching vector $b_\text{max}$, and depth $d$ is given by $(b_\text{max}^{d})^{N}$. Exploring all these possible configurations is computationally expensive, especially since the functions we want to calculate for these LTCs, like the rate of Eq. \eqref{rate_formal}, are recursive and resource-intensive. On top of this, we need to perform a maximization for the rate over a wide range of parameter values, further increasing the computational demand. To address this, we employ an ansatz that narrows the set of asymmetric trees under consideration. The procedure we follow to find this reduced set is the following.
Assume we want to find the optimal asymmetric tree-graph in terms of the rate for a certain communication distance $L$ and a fixed number of repeating stations $m$. We first find the best performing symmetric tree-graph, by running the optimization over all possible symmetric tree-graphs bounded by a certain $b_\text{max}$, and depth $d$. After identifying the optimal symmetric tree-graph, we introduce slight modifications to specific branches of it, to generate asymmetric tree-graphs. This involves incrementing or decrementing branch elements by small values (e.g., 1 or 2) in all possible ways, generating a diversified set.  
Then we can finally run the optimization for the rate over the so obtained reduced set of asymmetric tree-graphs.
For small tree-graphs (i.e. small values of $b_\text{max}$ and $d$), we observe that our method successfully spans the optimal asymmetric LTCs identified through exhaustive optimization across the full set of possibilities. This consistency suggests that this ansatz effectively captures the optimal solutions while significantly reducing the computational overhead.

\subsection{Optimal LTCs }
\begin{figure}
    \centering
    \includegraphics[width=1\linewidth]{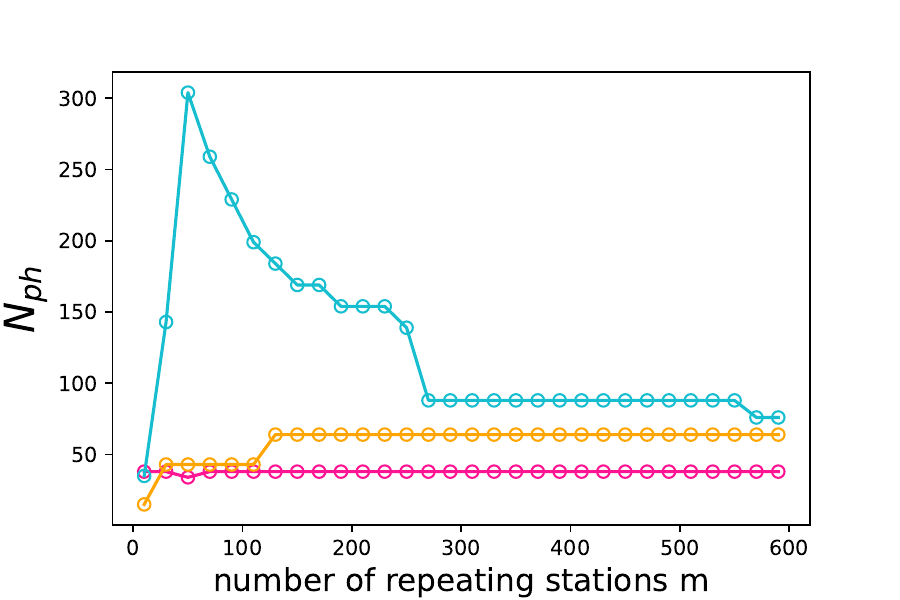}
    \caption{The number of photons $N_{ph}$ for the optimal tree-graphs of Fig. \ref{fig_1} (d) obtained maximizing the rate in terms of tree-graph geometry for a fixed communication distance $L = 300\text{km}$ across different numbers of repeating stations $m$. With {\protect\tikz \protect\draw[draw=GCyan, thick] (0,0) circle (0.3em) -- (-0.4,0) -- (0.4,0); } we denote the symmetric tree-graphs with maximum branching vector set to 20, with {\protect\tikz \protect\draw[draw=orange, thick] (0,0) circle (0.3em) -- (-0.4,0) -- (0.4,0); } the symmetric tree-graphs with maximum branching vector equal to 5 and with {\protect\tikz \protect\draw[draw=deeppink, thick] (0,0) circle (0.3em) -- (-0.4,0) -- (0.4,0); } the asymmetric trees generable with our protocol. }
    \label{N_ph fig 1d}
\end{figure}

\begin{figure}
    \centering
    \includegraphics[width=1\linewidth]{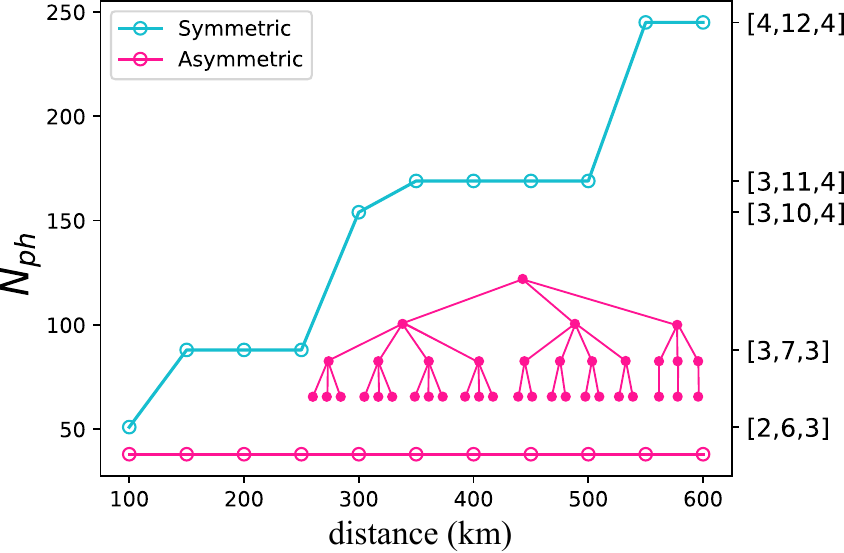}
    \caption{The optimal tree-graphs, obtained by maximizing the rate shown in Fig. \ref{fig_plots} (b), for every distance. The left y-axis the photon number, while the right y-axis lists the tree geometry for the symmetric case. For the asymmetric case the optimal tree-graph is the same for every distance and is represented in the figure. }
    \label{nph bis}
\end{figure}

We present here the optimal LTCs found for the optimizations across different parameters presented in the main text. In Fig.~\ref{N_ph fig 1d} we report the number of photons of the optimal tree-graphs obtained for the maximization of the rate shown in Fig.~\ref{fig_1}. The used parameters in this case are; $\eta_\text{d} = 0.95$, $L_\text{att} = 20\:\text{km}$ , $\epsilon_\text{r} = 10^{-4}$, $\tau_\text{ph} = 1\:\text{ns}$ and $\tau_\text{CZ} = 10 \tau_\text{ph}$. In tables~\ref{tab 1}, \ref{tab 2} and \ref{tab 3}, the specific geometry of these tree-graphs is indicated. In Fig.~\ref{nph bis}, on the right y-axes the geometries of the symmetric LTCs obtained in the optimization of Fig. \ref{fig_plots} (b) are specified. Also for the latter optimization the parameters appearing in the rate are the same used in the optimization of Fig.~\ref{fig_1}.\ref{tab success prob} lists the optimal symmetric and asymmetric LTCs obtained from the optimization of $\epsilon_\text{eff}$ shown in Fig.~\ref{fig_6_7} (a).  In Tab.~\ref{tab implementations} the rates displayed in Fig.~\ref{fig_6_7} (b) for different physical implementations, the optimal tree-graphs and the optimal number of repeating stations are reported.

\begin{table}[]
    \centering
    \begin{tabular}{|c|c|c|}
    \hline
    \textbf{tree-graph {\protect\tikz \protect\draw[draw=GCyan, thick] (0,0) circle (0.3em) -- (-0.4,0) -- (0.4,0); } } & \(\mathbf{N_{ph}}\) & \(\mathbf{m}\) \\
    \hline

    \([2,1, 15]\) & 35 & 10 \\ \hline
    \([2,14, 4]\) & 143 & 30 \\ \hline
    \([3,20, 4]\) & 304 & 50 \\ \hline
    \([3,17, 4]\) & 259 & 70 \\ \hline
    \([3,15, 4]\) & 229 & 90 \\ \hline
    \([3,13, 4]\) & 199 & 110 \\ \hline
    \([3,12, 4]\) & 184 & 130 \\ \hline
    \([3,11, 4]\) & 169 & 150-170 \\ \hline
    \([3,10, 4]\) & 154 & 190-230 \\ \hline
    \([3,9, 4]\) & 139 & 250 \\ \hline
    \([3,7, 3]\) & 88 & 270-550 \\ \hline
    \([3,6, 3]\) & 76 & 570-590 \\ \hline

    \end{tabular}
    
    \caption{ The optimal symmetric tree-graphs of the plot in Fig. \ref{fig_1}(d), obtained by maximizing the rate in terms of tree-graph geometry (with a maximum branching vector equal to 20) for a fixed communication distance $L = 300\text{km}$ across different numbers of repeating stations $m$. Each tree represents the best-performing symmetric configuration for the given fixed L and $m$.}
    \label{tab 1}
\end{table}
\begin{table}[]
    \centering
    \begin{tabular}{|c|c|c|}
    \hline
    \textbf{tree-graph {\protect\tikz \protect\draw[draw=orange, thick] (0,0) circle (0.3em) -- (-0.4,0) -- (0.4,0); } } & \(\mathbf{N_{ph}}\) & \(\mathbf{m}\) \\
    \hline
     \([2,1, 5]\) & 15 &10 \\ \hline
     \([2,5, 3]\) & 43 &30-110\\ \hline
     \([3,5,3]\) & 64 &130-590\\ \hline
    \end{tabular}
    \caption{The optimal symmetric tree-graphs of the plot in Fig. \ref{fig_1}(d), obtained by maximizing the rate in terms of tree-graph geometry (with a maximum branching vector equal to 5) for a fixed communication distance $L = 300\text{km}$ across different numbers of repeating stations $m$. Each tree represents the best-performing symmetric configuration for the given fixed L and $m$.}
    \label{tab 2}
\end{table}

\begin{table}[]
    \centering
    \begin{tabular}{|c|c|c|}
    \hline
    \textbf{tree-graph {\protect\tikz \protect\draw[draw=deeppink, thick] (0,0) circle (0.3em) -- (-0.4,0) -- (0.4,0); } } & \(\mathbf{N_{ph}}\) & \(\mathbf{m}\) \\
    \hline

    \([(4, 4), (4, 1), (3, 1)]\) & 38 &10-20\\ \hline
    \([(4, 3), (4, 1), (3, 1)]\) & 34 &30 \\ \hline
    \([(4, 3), (4, 2), (3, 1)]\) & 38 & 30-590\\ \hline

    \end{tabular}
    \caption{The optimal asymmetric tree-graphs of the plot in Fig. \ref{fig_1}(d), obtained maximizing the rate in terms of tree-graph geometry for a fixed communication distance $L = 300\text{km}$ across different numbers of repeating stations $m$. Each tree represents the best-performing asymmetric tree-graph, generable with our single emitter top-bottom generation, for the given fixed L and $m$. }
    \label{tab 3}
\end{table}

\begin{table}
\centering
\begin{tabular}{|c|c|c|}
\hline
\textbf{sym tree-graph {\protect\tikz \protect\draw[draw=GCyan, thick] (0,0) circle (0.3em) -- (-0.4,0) -- (0.4,0); }} & \textbf{asym tree-graph {\protect\tikz \protect\draw[draw=deeppink, thick] (0,0) circle (0.3em) -- (-0.4,0) -- (0.4,0); }} & \textbf{$\epsilon$} \\
\hline
\([4,5, 3]\) & \([(6, 4), (6, 4), (6, 3), (5, 1)]\) & 0.01 \\ \hline
\([4,5, 3]\) & \([(7, 4), (6, 3), (6, 2), (6, 2)]\) & 0.03 \\ \hline
\([4,5, 3]\) & \([(7, 4), (7, 3), (6, 2), (6, 1)]\) & 0.05 \\\hline
\([4,5, 3]\) & \([(7, 4), (7, 3), (7, 2), (5, 1)]\) & 0.07 \\\hline
\([3,7, 3]\) & \([(9, 4), (8, 3), (6, 2)]\) & 0.09-0.23 \\\hline
\([3,8, 3]\) & \([(10, 4), (9, 3), (5, 1)]\) & 0.25-0.27 \\\hline
\([3,8, 3]\) & \([(10, 4), (9, 3), (5, 1)]\) & 0.29-0.33 \\\hline
\([2,10, 3]\) & \([(10, 3), (10, 3)]\) & 0.35 \\\hline
\([2,10, 3]\) & \([(11, 4), (10, 3)]\) & 0.37-0.41 \\\hline
\([2,9, 3]\) & \([(11, 5), (10, 2)]\) & 0.43-0.45 \\\hline
\([2,9, 4]\)  & \([(9, 3), (9, 3)]\) & 0.47 \\\hline
\([2,9, 4]\)  & \([(10, 5), (11, 2)]\) & 0.49 \\\hline

\hline
\end{tabular}
\caption{
Optimal symmetric and asymmetric tree-graphs obtained from the optimization of $\epsilon_\text{eff}$, shown in the main plot of Fig.\ref{fig_6_7}(a). Each optimal tree is obtained by fixing a value of $\epsilon$ and determining the geometry that minimizes the value of $\epsilon_\text{eff}$.
 }
\label{tab success prob}
\end{table}

\begin{table}[]
    \centering
    \begin{tabular}{|c|c|c|c|}
    \hline
    \textbf{physical imp.} &\textbf{rate} & \textbf{tree-graph }& \textbf{m} \\
    \hline

    quantum dot \textcolor{tabgreen}{\ding{117}}&117.3 kHz & \([(4, 2), (4, 2), (4, 1)]\)  &44\\ \hline
    silicon vacancy \textcolor{blue}{\ding{117}}&3.9 kHz & \([(3, 2), (4, 2), (4, 1)]\) &57 \\ \hline
    neutral atom \textcolor{orange}{\ding{117}}& 273.9 Hz & \([(4, 2), (4, 2), (4, 1)]\)  & 57\\ \hline

    \end{tabular}
    \caption{We report the maximum rates reached, the optimal tree-graph used and the optimal number of repeating stations, for different physical implementations, over a communication distance of 100 km.\\ \textcolor{tabgreen}{\ding{117}} : $\gamma= 2\pi \times 100$ GHz \cite{gamma_qdot}, $t_{coh}= 4 $ $\mu $s \cite{tcoh_qdot} \\ \textcolor{blue}{\ding{117}} : $\gamma= 2\pi \times 2$ GHz \cite{bhaskar2020experimental}, $t_{coh}= 13 $ ms \cite{Silicon_vacancy_coherece}\\ \textcolor{orange}{\ding{117}} : $\gamma= 2\pi \times 170$ MHz \cite{gamma_atomo}, $t_{coh}= 1 $ s \cite{tcoh_atomo}}
    \label{tab implementations}
\end{table}
\subsection{Average photon number}
We now explain how the inset plot of Fig.\ref{fig_6_7} (a) is obtained. For a fixed loss probability ($\epsilon = 0.1$ in this case), we compute, for all symmetric trees, with $N_\text{ph} < 10^3 $ and maximum depth 3, the effective loss rate $\epsilon_\text{eff}$. For each value of $N_\text{ph}$ we select the tree-graph with the best recovery performance. The optimal tree-graphs are then grouped based on $\epsilon_\text{eff}$, where the range of $\epsilon_\text{eff}$ is divided in logarithmic intervals. Finally, within each group, we compute the average $N_\text{ph}$. The same procedure is made also for the asymmetric case, with the set of asymmetric tree-graphs obtained from the optimal symmetric ones as described in \ref{asymmetric generation}. \\
\section{Long range quantum communication}
Here, we present the results for the communication rate at high distances achieved using our protocol and an extended version of it, consisting in two quantum emitters arranged sequentially along the same waveguide. This modification expands the range of tree-graph geometries that can be generated top-bottom, enhancing the loss-tolerance capabilities and hence the rate in the case of long range quantum communication. Specifically, with this setup we can emit trees with eight branches, two of them can be of the form $\bm{b} = \{7,n\}$, and six of the form $\bm{b} = \{8,n\}$.
\begin{figure}
    \centering
    \includegraphics[width=1\linewidth]{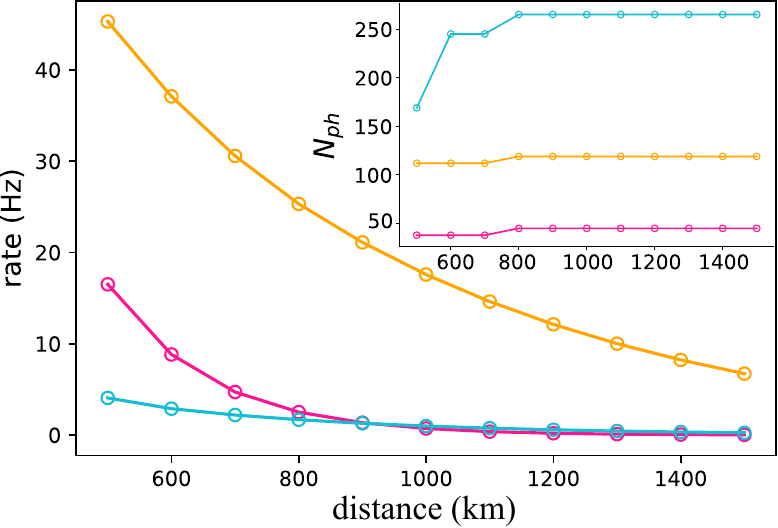}
    \caption{In the  main plot the maximized rate $\mathcal{R}$, for $m$ and the tree-graph geometry, in terms of the communication length $L$. The markers denote: {\protect\tikz \protect\draw[draw=deeppink, thick] (0,0) circle (0.3em) -- (-0.4,0) -- (0.4,0); } optimized rate obtained with our protocol with one emitter, {\protect\tikz \protect\draw[draw=orange, thick] (0,0) circle (0.3em) -- (-0.4,0) -- (0.4,0); } optimized rate obtained with our protocol with two emitters,{\protect\tikz \protect\draw[draw=GCyan, thick] (0,0) circle (0.3em) -- (-0.4,0) -- (0.4,0); } optimized rate obtained with the protocol of ref. \cite{HANNES_PRX}, taking into account symmetric trees of depth 3 with maximum  branching element equal to 20. The constants appearing in $\mathcal{R}$ are the same in all tree cases, and are; $\eta_\text{d} = 0.95$, $L_\text{att} = 20\: \text{km}$ , $\epsilon_\text{r} = 10^{-4}$, $\tau_\text{ph} = 1\:$ and $\tau_\text{CZ} = 10 \tau_{ph}$. The inset displays the number of photons of the optimized tree-graphs for the rate, for every distance.}
    \label{double emitter}
\end{figure}
Fig. \ref{double emitter} displays the rate as a function of the communication distance for the protocol using symmetric trees \cite{HANNES_PRX}, and our protocol with 1 and 2 emitter. In Tab. \ref{tab double emitter rate} we also report the optimized tree-graphs for every distance.
\begin{table}[h!]
\centering
\begin{tabular}{|c|c|}
\hline
\textbf{distance(km)} & \textbf{tree-graph} \\
\hline
500 & \begin{tabular}[c]{@{}l@{}}$[(4, 3),(4, 2),(3, 1)]$ \\ $[(8, 4),(8, 3),(7, 2),(7, 1)]$ \\ $[3,11, 4]$\end{tabular} \\
\hline
600--700 & \begin{tabular}[c]{@{}l@{}}$[(4, 3),(4, 2),(3, 1)]$ \\ $[(8, 4),(8, 3),(7, 2),(7, 1)]$ \\ $[4,12, 4]$\end{tabular} \\
\hline
800--1500 & \begin{tabular}[c]{@{}l@{}}$[(4, 3),(4, 2),(3, 1),(3, 1)]$ \\ $[(8, 4),(8, 3),(7, 2),(7, 2)]$ \\ $[4,13, 4]$\end{tabular} \\
\hline
\end{tabular}
\caption{The optimal tree-graphs of Fig. \ref{double emitter}, obtained by maximizing the rate for various communication distances $L$. We report the so obtained optimal tree-graphs generable with our protocol wit 1 and 2 emitters, and for the protocol of ref. \cite{HANNES_PRX} with maximum branching vector set to 20.}
\label{tab double emitter rate}
\end{table}
We see that the protocol utilizing top-bottom generation with two emitters ensures high communication rates, even at distances where the other two methods yield rates that are nearly zero.
In Fig. \ref{color 1000}, we show the rate as a function of the parameters $\gamma$ and $t_{coh}$ for a distance $L = 1000$ km, for our protocol with one and two emitters. With different markers, we highlight various technological implementations. Here, we also consider a quantum dot (\textcolor{GCyan}{\ding{117}}) with enhanced coherence time \cite{qdot_high_coherence}, which could offer a significant advantage in terms of rate, particularly in the context of long-range communication scenarios. Finally in Tab.\ref{tab: color plots} the rate, the optimal tree-graphs and the optimal number of repeating stations for the two cases are presented.
\begin{figure*}[t!]
    \begin{minipage}{1\textwidth}
    \includegraphics[width=\textwidth]{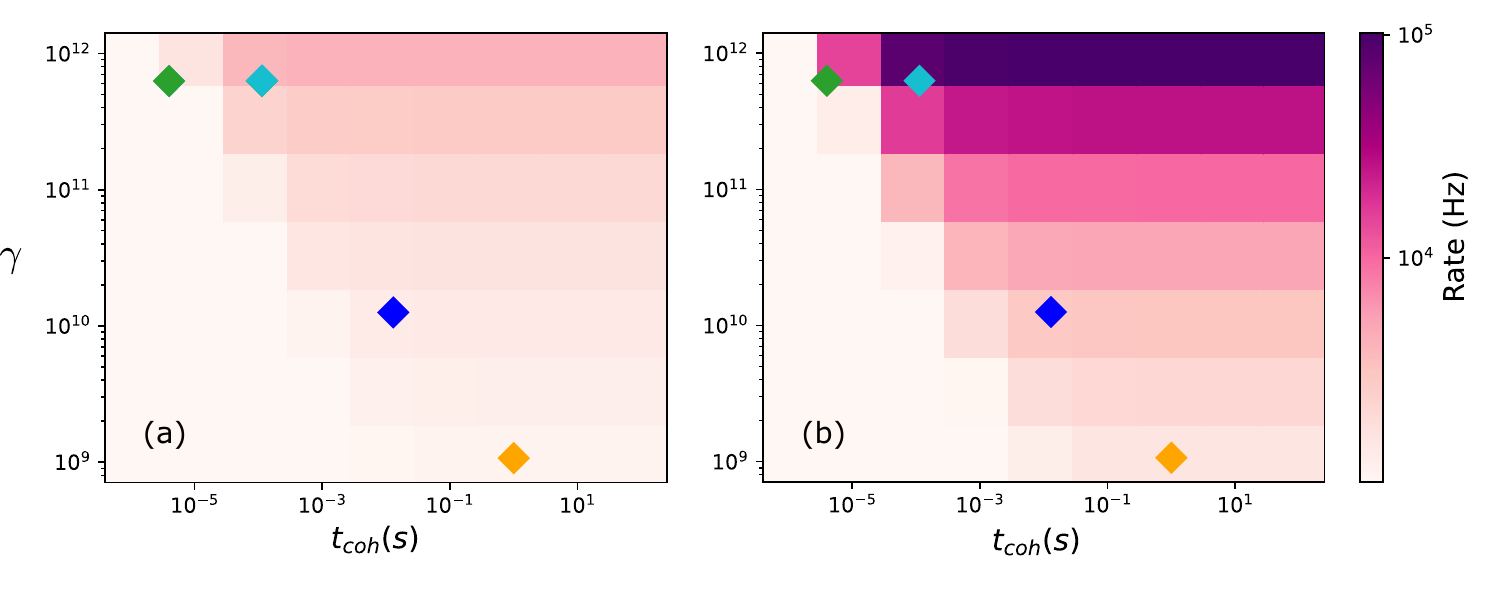}
    \end{minipage}%
\caption{ The rate $\mathcal{R}$ in function of the optical line width $\gamma$ and the coherence time of the quantum emitter $t_\text{coh}$, for $L = 1000 $ km is reported, for (a) our protocol with a single emitter (b) our protocol with two emitters. The markers correspond to different technological implementations for the emitters. \textcolor{tabgreen}{\ding{117}} a single semiconductor quantum dot coupled with a nano-cavity,  \textcolor{GCyan}{\ding{117}} a single semiconductor quantum dot coupled with a nano-cavity with enhanced coherence time \textcolor{blue}{\ding{117}} a single silicon-vacancy (SiV) color center coupled with a photonic crystal cavity and \textcolor{orange}{\ding{117}} a single neutral atom coupled
with a fiber Fabry-Perot cavity.}
\label{color 1000}
\end{figure*}
\begin{table}[ht]
    \centering
    \textbf{Single emitter} 

    \vspace{0.3cm} 

    \begin{tabular}{|c|c|c|c|}
    \hline
    \textbf{physical implementation} & \textbf{rate} & \textbf{tree-graph }& \textbf{m} \\
     \hline
    quantum dot \textcolor{GCyan}{\ding{117}} & 3.5 kHz & \([(4, 3), (4, 2), (3, 1), (3, 1)]\) & 1000 \\
    \hline

    quantum dot \textcolor{tabgreen}{\ding{117}} & 0 kHz & \(-\) & - \\\hline
    silicon vacancy \textcolor{blue}{\ding{117}} & 76.1 Hz & \([(4, 3), (4, 2), (3, 1), (3, 1)]\) & 1000 \\\hline
    neutral atom \textcolor{orange}{\ding{117}} & 1.7 Hz & \([(4, 3), (4, 2), (3, 1)]\) & 1000 \\\hline
   
    \end{tabular}
    
    \vspace{0.7cm} 

    \textbf{Double emitter} 

    \vspace{0.3cm} 

    \begin{tabular}{|c|c|c|c|}
    \hline
    \textbf{physical implementation} & \textbf{rate} & \textbf{tree-graph }& \textbf{m} \\

    \hline

    quantum dot \textcolor{GCyan}{\ding{117}} & 26.1 kHz & \([(7, 4), (8, 3), (8, 2), (7, 1)]\) & 805 \\
    \hline

    quantum dot \textcolor{tabgreen}{\ding{117}} & 0 kHz& \(-\) & - \\\hline
    silicon vacancy \textcolor{blue}{\ding{117}} &  582.7 Hz & \([(7, 4), (8, 3), (8, 2), (7, 1)]\) & 855 \\\hline
    neutral atom \textcolor{orange}{\ding{117}} & 34.3 Hz & \([(7, 4), (8, 3), (8, 2), (7, 2)]\) & 1000 \\\hline
    \end{tabular}
    \caption{We report the maximum rates reached, the optimal used tree-graph, and the optimal number of repeating stations, for different physical implementations, over a communication distance of 1000 km, for the single, and the double emitter protocol. \\
     \textcolor{GCyan}{\ding{117}} : $\gamma= 2\pi \times 100$ GHz \cite{gamma_qdot}, $t_{coh}= 113 $ $\mu $s \cite{qdot_high_coherence} \\
    \textcolor{tabgreen}{\ding{117}} : $\gamma= 2\pi \times 100$ GHz, $t_{coh}= 4 $ $\mu $s \cite{tcoh_qdot}\\
    \textcolor{blue}{\ding{117}} : $\gamma= 2\pi \times 2$ GHz \cite{bhaskar2020experimental}, $t_{coh}= 13 $ ms \cite{Silicon_vacancy_coherece}\\
    \textcolor{orange}{\ding{117}} : $\gamma= 2\pi \times 170$ MHz \cite{gamma_atomo}, $t_{coh}= 1 $ s \cite{tcoh_atomo} }
    \label{tab: color plots}
\end{table}
\section{Correcting qubit errors with majority vote strategies}\label{error model}
Here, we study more in depth the impact of errors in our generation protocol. For this, we now employ a more detailed error model, originally introduced for symmetric tree-graph LTCs~\cite{ECONOMOU_PERFORMANCE, azuma2015all}, and we extend it to asymmetric codes. Similarly to Refs.~\cite{ECONOMOU_PERFORMANCE, azuma2015all}, we also incorporate a majority vote strategy: this makes explicit use of the abundant redundancy in the tree-graph encoding, to also correct qubit errors (that is, errors which are not losses).\\
\indent The secret key fraction in Eq.~\eqref{secret key fraction} is not only affected by losses, but also by errors in the photonic qubits. Specifically, we assume
that each photon undergoes an independent depolarization channel, described as,
\begin{equation}
\Delta_\mu(\rho) = (1 - \mu) \rho + \frac{\mu}{3} \left( \mathrm{X} \rho \mathrm{X} + \mathrm{Y} \rho \mathrm{Y} + \mathrm{Z} \rho \mathrm{Z} \right),
\end{equation}
where $\rho$ is the density matrix of a single photon. The probability of an error for a single photon measurement under any basis is then $\mu_\text{sp} = \frac{2}{3}\mu$. We are now interested in the dependence of the transmission fidelity $F$ on the parameter $\mu_\text{sp}$ in the case of asymmetric trees. The fidelity can be expressed as  
\begin{equation}
F = (1 - \bar{e}_{\mathrm{decoding}})^{m+1},
\end{equation}
where $m$ is the number of repeater stations and $\bar{e}_{\mathrm{decoding}}$ is the error probability of having a logical error in a decoded qubit conditioned on the fact that all adaptive measurements in the recovery protocol apparently succeeds (i.e., we are able to consistently complete the decoding-encoding), but the output is actually incorrect due to errors in the measurement process; the latter is given by
\begin{equation}
\bar{e}_{\mathrm{decoding}} = \frac{\bar{e}_{\mathrm{incorrect}}}{P_{\text{s}}}.
\end{equation}
Here, $P_\text{s}$ is the probability of successfully performing all the adaptive measurements, as given in Eq. \ref{Ps genrale 3} for asymmetric graphs, and $\bar{e}_{\mathrm{incorrect}}$ is the probability of decoding an incorrect result at a repeater node. To determine $\bar{e}_{\mathrm{incorrect}}$,
it is useful to remember that to successfully decode a result (either correct or incorrect) at each repeater node, one photon in the first level of the tree needs to be measured in the $\mathrm{X}$ basis, whereas all its leaf photons in the second level, together with all other photons in the first level, need to be measured in the $\mathrm{Z}$ basis, either directly or indirectly. Hence, correct decoding requires three conditions to be fulfilled: (1) The first level photon measured in the $\mathrm{X}$ basis must be errorless;  (2) the parity of all $\mathrm{Z}$ measurement outcomes from the remaining first-level photons, whether measured directly or indirectly, must be errorless; (3) The parity of all $\mathrm{Z}$ measurement outcomes from the second-level photons connected to the first-level photon, measured in the $\mathrm{X}$ basis, either directly or indirectly, must also be errorless.\\
\indent In the following, let $l$ be the number of photons in the first level which are lost and need to be measured indirectly, $n$ the number of photons which are not lost and are measured directly, and the remaining photons that are not lost but can be measured indirectly are labeled as $b'= b_0-n-l $. We now account for all possible scenarios. i.e. all possible combinations of $n$,$l$ and $b'$. For symmetric graphs, this can be achieved as discussed in Refs.~\cite{ECONOMOU_PERFORMANCE, azuma2015all}. In the asymmetric case, we have to explicitly compute different permutation scenarios, because in general every branch of the tree has a different geometry. Thus, $\bar{e}_{\mathrm{incorrect}}$ in the case of asymmetric tree-graphs reads: 
\begin{widetext}
\begin{align}
\bar{e}_{\mathrm{incorrect}} &= 
\sum_{l=0}^{b_0-1} \sum_{n=0}^{b_0 - l} 
\prod_{p \in \mathcal{P}(b_0,l,n)}
\Bigg[
\prod_{l'\in p } \epsilon R_0^{(l')} 
\prod_{n' \in p} (1-\epsilon)R_0^{(n')} 
\prod_{b'\in p}(1-\epsilon)R_0^{(b')}
\Bigg] \nonumber \\
&\times 
\sum_{q=0}^{b_0} \frac{\epsilon^{q}(1-\epsilon)}{(1-\epsilon^{b_0})} 
\sum_{m=0}^{b_0^{(q)}}
\left[(1-\epsilon)(1-R_1^{(q)})\right]^m 
\left[R_1^{(m)}\right]^{b_0^{(q)}-m} 
\bar{e}_{n, m},
\end{align}
\end{widetext}
where $\mathcal{P}(b_0,l,n)$ indicates the set of all possible combinations of the $b_0$ indices of an asymmetric tree-graph (each index labeling a different branch), divided in three groups; $l'$ with $l$ elements, $n'$ with $n$ elements, and $b'$ with $b_0 - l - n $ elements. Let now $ R_n^{(i)} $ be the probability of successfully performing an indirect $Z$ measurement, on the $n$-th level of the $i$-th branch, and denote by $\bar{e}_{n, m}$ the error probability of the decoded qubit, in the case where $n$ of the $Z$ measurements in the first level, and $m$ of the second level, are performed through direct measurement, while all the others are performed indirectly.
Then, one has
\begin{equation}
    \bar{e}_{n, m} = 1-(1-\mu_{\text{sp}})\left( 1- \Tilde{P_n} \right)\left( 1- \Tilde{P}_m(q) \right),
\end{equation}
where we defined
\begin{widetext}
\begin{align}
    \Tilde{P}_n &= \sum_{i=0}^n\left[ \binom{n}{i} (1-\mu_{\text{sp}})^{n-i}\sum^{b_0 -1-n}_{\substack{j = 0 \\ \text{\text{mod}2[i+j]=1}}} \sum_{k,k' \in \mathcal{B}(b_0 -1-n,j)} \prod_{k} \bar{e}_{I_1}^{(k)} \prod_{k'} (1-\bar{e}_{I_1}^{(k')} )\right] \\
    \Tilde{P}_m(q) &= \sum_{i=0}^m \left[\binom{m}{i} \mu^i_{\text{sp}}(1-\mu_{\text{sp}})^{m-i} \sum^{b_0^{(q)}-m}_{\substack{j = 0 \\ \text{\text{mod}2[i+j]=1}}} \binom{b_0^{(q)}-m}{j}\bar{e}_{I_2}^{(q)}\right],
\end{align}
\end{widetext}
where $k,k'$ are sets of indexes (of the branches) of $j$ and $b_0-1-n-j$ elements, and $\mathcal{B}(b_0 -1-n,j)$ contains all the possible combinations of groups of indexes with that length. $\bar{e}_{I_s}^{(q)}$ is the error probability of guessing a wrong $Z$ measurement outcome on a level $s$ photon of the $q$-th branch, by doing an indirect measurement using the majority vote strategy. Since our asymmetric tree is made by branches that are symmetric trees, the expression for $\bar{e}_{I_s}^{(q)}$, is the one derived in the supplementary material of \cite{azuma2015all}. To relate $\mu_\text{sp}$ to experimental parameters, we denote by $\mu_\text{dep}$ the depolarization rate, and set
\begin{equation}
   \mu_{\mathrm{decoh}} = \frac{3}{4} \left[1 - \exp\left( \frac{-T_{\mathrm{graph}}}{t_{\mathrm{coh}} N_{\mathrm{ph}}} \right) \right]. 
\end{equation}
We then finally estimate the resulting depolarization rate
\begin{equation}
    \mu_{\mathrm{sp}} = \frac{2}{3} \left( \mu_{\mathrm{decoh}} + \mu_{\mathrm{dep}} - \frac{4}{3} \mu_{\mathrm{decoh}} \mu_{\mathrm{dep}} \right).
\end{equation}

\section{Emitter faults and correlated errors}
 Finally, we also include a more detailed analysis of emitter errors, which can potentially result in correlated errors across the LTC, which would potentially be beyond the correction capabilities of the majority vote strategy; by considering a worst case scenario (where any correlated error completely corrupts the transmission fidelity), we show that for most of the relevant physical implementations our protocol features remarkable resilience against such correlated errors. Specifically, we assume that any single error occurring during the generation of a LTC leads to an uncorrectable correlated error. Considering a scenario with  $m$ repeating stations, the probability of having at least one error in the generation of LTCs across the entire repeater chain is
\begin{equation}
P_\text{error} = 1-\left(e^{\frac{-T_{\text{graph}}}{t_\text{coh}}}\right)^m.
\end{equation} 
The mean fidelity can then be estimated as 
\begin{equation}
    F = (1-P_\text{error})F_0 + \frac{1}{2}P_\text{error},
\end{equation}
where $F_0$ is the fidelity discussed so far, which comprises both losses and uncorrelated Pauli errors, the latter being corrected though the majority vote strategy discussed in Section~\ref{error model}. 
\begin{figure}
    \centering
    \includegraphics[width=1\linewidth]{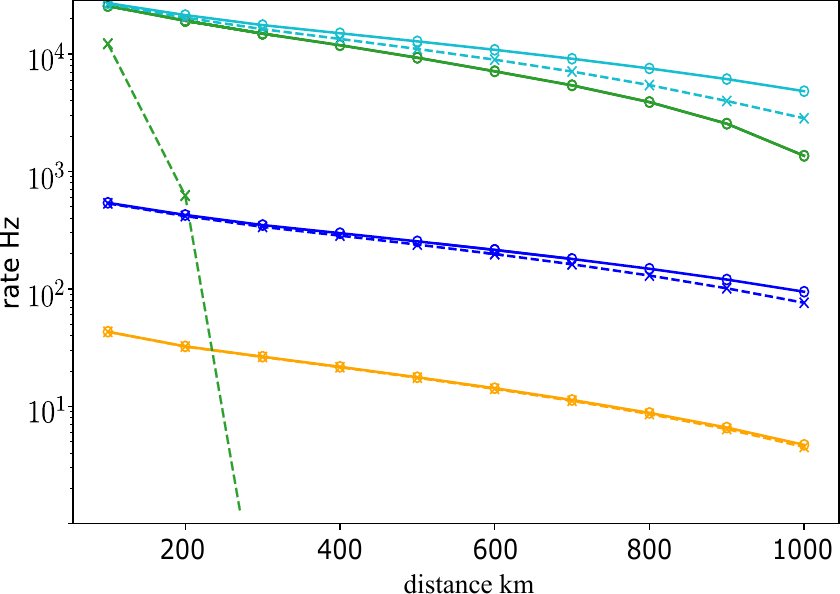}
    \caption{The rate for our top bottom single emitter protocol, optimized for different distances, over asymmetric tree geometries. The solid lines represent the rate considering the single Pauli error model and error correction based on the majority vote strategy. The dashed lines consider also the possibility of correlated errors, assuming a worst case scenario where every error in the generation of the tree cluster leads to a correlated error that is uncorrectable. Different colors correspond to different hardware implementations.\\
    \textit{Quantum dot} \textcolor{GCyan}{\ding{117}} : $\gamma= 2\pi \times 100$ GHz \cite{gamma_qdot}, $t_{coh}= 113 $ $\mu $s \cite{qdot_high_coherence} \\
    \textit{Quantum dot} \textcolor{tabgreen}{\ding{117}} : $\gamma= 2\pi \times 100$ GHz, $t_{coh}= 4 $ $\mu $s \cite{tcoh_qdot}\\
    \textit{Silicon vacancy} \textcolor{blue}{\ding{117}} : $\gamma= 2\pi \times 2$ GHz \cite{bhaskar2020experimental}, $t_{coh}= 13 $ ms \cite{Silicon_vacancy_coherece}\\
    \textit{Neutral atom} \textcolor{orange}{\ding{117}} : $\gamma= 2\pi \times 170$ MHz \cite{gamma_atomo}, $t_{coh}= 1 $ s \cite{tcoh_atomo} }
    \label{tab: color plots}
    
    \label{fig:worst case}
\end{figure}

In Fig.~\ref{fig:worst case} we report the optimized rate of Eq.~\ref{rate_formal} as a function of distance, for different physical platforms and under two distinct error models. Solid lines correspond to the case where only single-qubit Pauli errors are considered, while dashed lines include the effect of correlated errors during the generation, assuming a worst case scenario as described above. We find that the impact of correlated errors from emitter faults depends heavily on the specific platform considered, with the most detrimental effects affecting systems where the decoherence rate is fast with respect to the operation rate at the repeater stations. On the other hand, systems where the generation of the LTCs is fast enough with respect to the error rate of the emitter do not suffer tremendously of such faults. More specifically, our simulations show that for platforms such as neutral atoms and vacancy centers in diamond, where the emitter coherence time is sufficiently long compared to the emission time, the effect of correlated errors is negligible, even in the pessimistic worst-case model that we considered. In contrast, quantum dots, which typically have shorter coherence times, appear to be much more sensitive to such errors.\\
\indent In conclusion, under correlated errors the ratio between the emission time and the coherence time of the emitter plays a critical role. In the case of neutral atoms, where the coherence time is significantly longer than the emission time, the impact of correlated errors is virtually negligible. In contrast, quantum dots, which typically exhibit shorter coherence times, appear to be highly susceptible to this type of error.
We stress that in practice not every error during generation would result in a correlated and uncorrectable error, and that in principle, even correlated errors can be corrected depending on their structure. Moreover, it is worth noting that modest improvements in the coherence time can be sufficient to shift the system into a regime where it becomes resilient to correlated errors. This is illustrated by the cyan curve in Fig.~\ref{fig:worst case}, which corresponds to a quantum dot with enhanced coherence time.


\newpage

\bibliographystyle{apsrev4-2}
\bibliography{biblio_updated}

\end{document}